\documentclass[aps,prl,reprint,preprintnumbers,amsmath,amssymb,floatfix]{revtex4-1}

\usepackage{longtable}
\usepackage{morefloats}
\usepackage[dvips]{graphicx}
\usepackage{color}
\usepackage{epsfig,graphicx,amsfonts,amsbsy}
\usepackage{amsmath,amsfonts,amsthm,amssymb}
\usepackage{appendix}
\usepackage{bbm}
\usepackage{makeidx}
\usepackage{url}
\usepackage{verbatim}
\usepackage[bookmarksnumbered,pdfpagelabels=true,plainpages=false,colorlinks=true,linkcolor=blue,citecolor=blue,urlcolor=blue]{hyperref}
\usepackage[rightcaption]{sidecap}
\usepackage{array}
\usepackage{booktabs}
\usepackage{multirow}
\usepackage{bbm}
\usepackage{tabularx}
\usepackage{cancel,soul,ulem}

\newcommand{\bs}[1]{\boldsymbol{#1}}

\makeatletter

\begin{document}
\title{Fingerprints of Universal Spin-Stiffness Jump in Two-Dimensional Ferromagnets}
\author{Roberto E. Troncoso}
\author{Arne Brataas}
\author{Asle Sudb\o}
\affiliation{Center for Quantum Spintronics, Department of Physics, Norwegian University of Science and Technology, NO-7491 Trondheim, Norway}

\begin{abstract}
Motivated by recent progress on synthesizing 
two-dimensional magnetic van der Waals systems, we propose a setup for detecting  the topological Berezinskii-Kosterlitz-Thouless (BKT) phase transition in spin-transport experiments on such structures. We demonstrate that the spatial correlations of injected spin-currents into a pair of metallic leads
can be used to measure the predicted universal jump of $2/\pi$ in the ferromagnet spin-stiffness as well as its predicted universal square root dependence on temperature as the transition is approached from below. 
Our setup provides a simple route to measuring this topological phase transition in two-dimensional magnetic systems, something which up to now has proven elusive. It is hoped that this will encourage experimental efforts to investigate critical phenomena beyond the standard Ginzburg-Landau paradigm in low-dimensional magnetic systems with no local order parameter.
\end{abstract}

\maketitle
{\it Introduction}.--Phase transitions of matter are  ubiquitous and fascinating phenomena in nature. In particular, transitions occurring in two-dimensions (2D) have long attracted great interest because of their intriguing physics. Quantum or thermal fluctuations play a fundamental role in the stability of phases. The special features of low dimensional systems with continuous symmetries, are rooted in the celebrated Hohenberg-Mermin-Wagner theorem \cite{Mermin-Wagner,Hohenberg1967,Coleman1973} stating that there can be no spontaneous breaking of a continuous symmetry at any finite temperature in dimensions $d\le 2$. This prevents the existence of long-range order at any temperature  $T > 0$, resulting in the absence of a local order parameter. Nonetheless, a low-temperature phase free of {\it topological defects} can exist, featuring quasi long-range order characterized by algebraically decaying correlations. 

Phase transitions may quite generally be thought of as mediated by the proliferation of topological defects with a concomitant loss of some generalized stiffness \cite{Anderson}. Unfortunately, in most cases it is almost impossible to make these notions precise and quantitative, at least analytically.  
However, the precise mechanism by which this happens in low-dimensional magnets, superfluids, and crystals has been elucidated in a series of seminal works of Berezinski\v{i}, Kosterlitz, and Thouless (BKT) \cite{Berezinskii,Burch2018,Kosterlitz1973,Kosterlitz_1974}, describing a topological phase transition at the critical temperature $T_{{c}}$ from tightly bound pairs of vortices and anti-vortices to an unpaired disordered phase. 

A remarkable and unique prediction of this theory is a specific feature characterizing the vanishing of topological order, or stiffness, as the system approaches the critical temperature from below \cite{Nelson_Kosterlitz}. This is expressed by the long-wavelength relation
\begin{align}\label{eq: universaljump}
	\frac{1}{{\cal K}_R(T)}=\frac{1}{{\cal K}}+4\pi^2\lim_{q\rightarrow 0}\frac{\langle {n}_{q}{n}_{-q}\rangle}{q^2},
\end{align}
where ${\cal K}$ is the stiffness of system, e.g., superfluid density or spin-stiffness in easy-plane ferromagnets, ${\cal K}_R$ is the vortex-renormalized stiffness and $n_q$ is the vorticity in momentum space. When the system reaches the transition temperature, the stiffness ${\cal K}_R$ jumps discontinuously to zero. The key feature of the transition is that this jump is 
{\it universal}, ${\cal K}_R\left(T_{{c}}\right)/k_BT_{{c}}=2/\pi$. 
This  was first verified experimentally in thin films of ${}^4$He \cite{Bishop1978} and later in other systems such as superconductors \cite{Gubser1979,Hebard1980,Voss1980,Wolf1981,Epstein1981}, colloidal crystals \cite{Halperin1978,Young1979,Zahn1999}, Josephson-junction arrays \cite{Resnick1981,Voss1982},  and ultracold atomic Bose gases \cite{Hadzibabic2006}. 

Despite its great interest and recent efforts \cite{bedoyapinto2020}, the observation of BKT transitions has proven elusive in spin systems, mainly due to the difficulty of manufacturing two-dimensional magnets. The discovery of graphene in 2004 \cite{Novoselov2004} was a turning point for significant experimental progress in fabricating atomically thin magnetic films, also known as 2D magnetic van der Waals (vdW) materials \cite{Huang2017,Burch2018,Deng2018,Klein2018,Bonilla2018,Puthirath2018,Song2018,Huang2018,Gibertini2019}.
\begin{figure}[h!]
	\includegraphics[width=\linewidth,clip=]{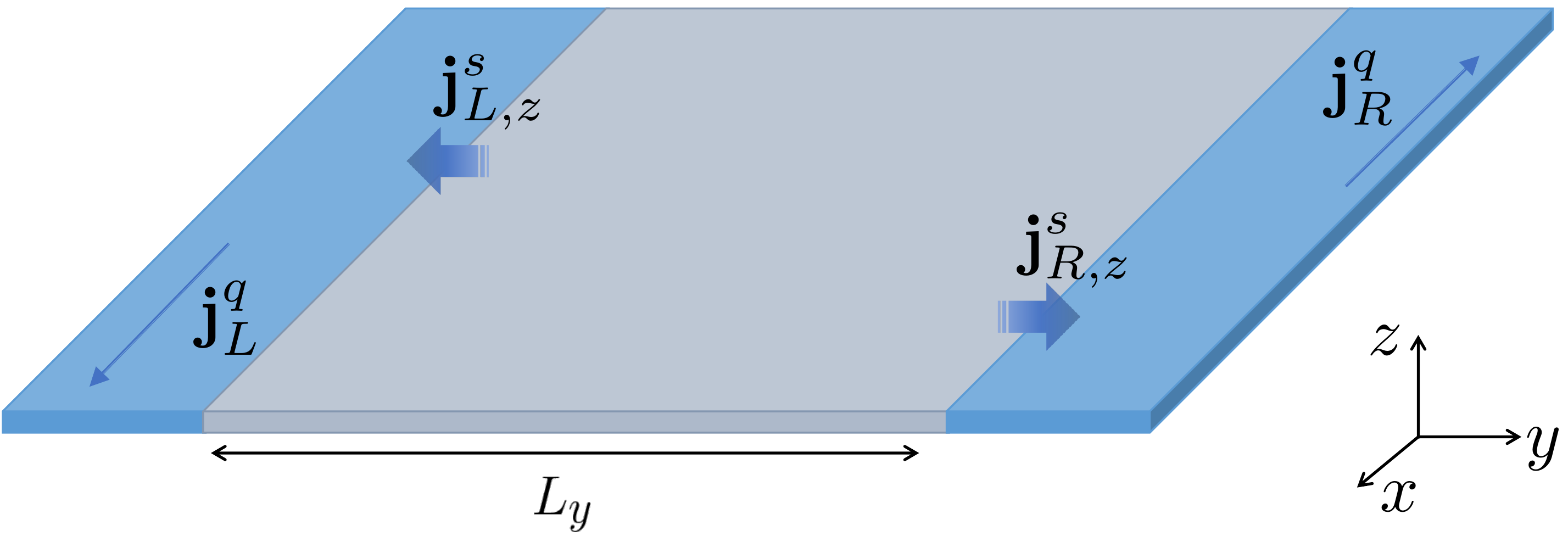}
	\caption{Schematic plot for the direct observation of BKT transition. The setup consists of a 2D magnetic vdW material attached at its ends to a couple of identical non-magnetic metallic leads. The detection is based on the measurement of charge-current cross-correlations between the left and right lead of the device geometry.}
	\label{fig:2D-setup}
\end{figure}

In this Letter, we propose an experimental setup for the observation of the BKT transition in 2D magnetic vdW materials using standard methods in the study of spin transport. It is based on electrical measurements of {charge-current cross-correlations} in metal-magnet hybrid structures \cite{Goennenwein2015,Cornelissen2015,lebrun2018,Kim2020}. In this approach, distant metallic leads detect spin-currents flowing into and out of the magnetic material, as displayed in Fig. \ref{fig:2D-setup}. At the center of this proposal is the phenomenon of spin and charge conversion, an inherent property of materials having strong spin-orbit coupling manifested in the spin-Hall effect (SHE) \cite{Sinova2015}. The characteristic spin dynamics in the magnetic insulator pump spin-currents into the normal metal \cite{Tserkovnyak2005} resulting in a spin accumulation that, in turn, induces charge-currents by the inverse spin-Hall effect (ISHE). The electrical detection of spin-currents signals, and their correlations, will provide direct access to thermally induced vortex proliferation and quasi-ordered -- disordered phase transition. This work constitutes a first step to studying low-dimensional phenomena involving notions relevant to  phase-transitions, 2D magnetic material science, and spintronics.

Nonlocal spin transport measurements in metal-magnet-metal heterostructures have proven useful for studying transport properties of pure spin-currents \cite{Goennenwein2015}, long-distance spin-transport \cite{Cornelissen2015,lebrun2018,Oyanagi2019}, magnon-polaron transport \cite{Cornelissen2017}, and viscosity in magnons systems \cite{Ulloa2019}. Recently, Ref. \cite{Bender2019} proposed a method of detecting the degree of coherence of magnon states by measuring spin-current correlations. This detection scheme is particularly useful since the spin-current cross-correlation is related to the power spectral density. Although the ideas developed in Ref. \cite{Bender2019} are general, the focus was on spatially homogeneous magnetization dynamics, e.g., ferromagnetic resonance and thermally induced magnons. In our approach, we tailor the detection scheme to a system that lacks long-range order. We show that spatially-dependent {charge-current cross-correlations} allow characterizations of the BKT phase transition. 

Our findings give direct access to track the evolution from algebraic- to exponential-decaying spin-spin correlations as the temperature increases. In particular, we provide an accessible route to measuring the universal jump of the spin-stiffness in 2D magnets as the temperature approaches the critical temperature. Taking advantage of the spin to charge conversion, this enables a clear-cut experimental demonstration in a magnetic system of one of the remarkable predictions of the BKT theory.

{\it Charge-current cross-correlation}.-- In the proposed setup, we consider a two-dimensional magnet, with dimensions $L_x$ and $L_y$, coupled at its ends to a couple of identical non-magnetic metallic leads, see Fig. \ref{fig:2D-setup}. Pure spin-currents into the left and right normal metals are related to electric signals via the inverse spin Hall effect \cite{Ando2011,Sinova2015}. We introduce the charge-current cross-correlation, defined by ${\cal C}^{(2)}(\tau)\equiv\langle I_{L}({\bs r},t)I_{R}({\bs r},t+\tau)\rangle$, where $\langle ,\rangle$ denotes a statistical average. Here, $I_{\ell}$, with $\ell=L, R$, is the net charge-current existing in the $\ell$ metallic lead. We are interested on the static correlation and thus in the evaluation of ${\cal C}^{(2)}\equiv {\cal C}^{(2)}(0)$. In the following, we relate ${\cal C}^{(2)}$ to the cross-correlation of the injected spin-currents using spin and charge conversion at the leads. This suffices to establish the connection between the characteristic spin-spin correlations and the universal behavior predicted in the BKT theory.

{\it Berezinski\v{i}-Kosterlitz-Thouless Transition}.-- We consider a ferromagnetic insulator with axially symmetric exchange coupling around the $\hat{\bs z}$ direction. The nearest-neighbor Hamiltonian is, $H=\sum_{\langle ij \rangle} J_{\alpha\beta}S_{\alpha i}S_{\beta j}$, with $i,j$ the position of spins and $\alpha,\beta=x,y,z$ labeling their components. The ferromagnetic exchange coupling is $J_{\alpha\beta}=-J_{\alpha}\delta_{\alpha\beta}$, with $J_x=J_y\equiv J$ and $J_z<J$, favoring spin ordering in the xy-plane.  Thus, we can express the normalized spin variable in terms of an angle $\theta_i$ on each lattice site $i$, as ${\bs S}_i=\left(\cos\theta_i,\sin\theta_i,0\right)^T$. The resulting Hamiltonian, known as the classical XY-model \cite{Kosterlitz2016}, is given by 
\begin{align}\label{eq:xymodel}
H=-J\sum_{\langle ij \rangle} \cos\left(\theta_i-\theta_j\right),
\end{align}
featuring a continuous $SO(2)$ or $U(1)$ symmetry. The XY-model \eqref{eq:xymodel} is also a useful model for superfluid helium \cite{Bishop1978,NELSON1979} and hexatic liquid crystals \cite{Chaikin1995}. At low temperatures, we take into account only small spin fluctuations. Thus, in the continuum approximation, smooth phase variations are described by the Hamiltonian $H={J}\int d{\bs r}\left(\nabla\theta\right)^2/{2}$. In two dimensions in the entire low-temperature {\it phase}, the spin-spin correlation function, defined by the statistical average $G({\bs r}-{\bs r}')=\langle {\bs S}({\bs r})\cdot{\bs S}({\bs r}')\rangle$.
decays algebraically, $1/\left|{\bs r}-{\bs r}'\right|^{\eta}$, with $\eta=k_BT/2\pi J$, indicative of critical behavior. Here, the critical exponent $\eta$ is the anomalous scaling dimension of the spin-field. At long distances, the spin correlations vanish, $\lim_{|{\bs r}-{\bs r}'|\rightarrow\infty} G({\bs r}-{\bs r}')=0$, corresponding to the absence of long-range order in accordance with the Mermin-Wagner theorem \cite{Mermin-Wagner}. Thus, spin-waves suffice to destroy long-range order. The entire low temperature phase is critical with an infinite correlation length and algebraic decay of spin-correlations.

In the high temperature phase, smooth spatial variations in $\theta({\bs r})$ no longer suffice to accurately describe the fluctuations.  Fluctuations beyond spin-waves are included by separating the vector $\nabla\theta=\nabla\theta_L+\nabla\theta_T$ into a longitudinal and a transverse part, the spin-waves and non-smooth variations respectively, defined by ${\nabla}\cdot\nabla\theta_T=0$ and ${\nabla}\times\nabla\theta_L=0$. Thus, the model is generalized to $H=\frac{J}{2}\int d{\bs r}\left[\left(\nabla\theta_L\right)^2+\left(\nabla\theta_T\right)^2\right]$, with the first term corresponding to the spin-wave part discussed above. The non-smooth variations in $\theta$ describe vortices, topological defects associated with the first homotopy group \cite{Nakahara} 
$\pi_1(U(1)) = \mathbb{Z}$ of the $2D$XY-model that have circulation $\oint d\vec{l}\cdot\nabla\theta_T=2\pi n$ with ''topological charge'' $n\in \mathbb{Z}$. At high temperature, the topological defects become important, and their statistical mechanics may be mapped to that of a $2D$ Coulomb gas with overall charge neutrality. The low-temperature phase where vortices and anti-vortices are tightly bound together is equivalent to an insulating dielectric state while the high temperature temperature phase with dissociated pairs of vortex$-$ anti-vortex pairs corresponds to a metallic phase. This transition is a topological phase-transition. The 
temperature at which it takes place is found by noting that the energy of a vortex-antivortex pair separated by a distance $r$, is $E=\pi J \ln\left(r/a\right)$ and the entropy $S=2\ln\left(r/a\right)$, where $a$ is some short-distance cutoff. A simple estimate for the Helmholtz free energy yield $F=\left(\pi J- 2T\right)\ln \left(r/a\right)$. It is clear that is energetically favorable to have free vortices when $T>T_{{c}}=\pi J/2k_B$. This estimate of $T_c$ ignores screening of two test-charges, and the description also ignores coupling between vortices and spin-waves. Taking such effects onto account slightly reduces $T_c$ without altering the universality class of the transition, which may be viewed as a special class of phase transition where the conformal invariance of standard critical points is lost, a ''conformality lost'' phase transition \cite{PhysRevD.80.125005,PhysRevD.100.085005}. In modern terms, the precise mechanism for this loss of conformality is the annihilation of an ultraviolet and and infrared fixed point as some marginal operator of the system is varied \cite{PhysRevD.80.125005,PhysRevD.100.085005}.   

In general, a phase transition may be viewed as a proliferation of topological defects characterised by an appropriate homotopy group with a concomitant loss of an associated generalised stiffness. In the present context, thermally induced spin-fluctuations (angle-fluctuations) reduce and eventually destroy the existing topological order. The generalised stiffness for our system, the spin-stiffness  ${\cal K}_R(T)$, is a global order-parameter for topological order. 

\begin{figure}[h!]
	\includegraphics[width=\linewidth,clip=]{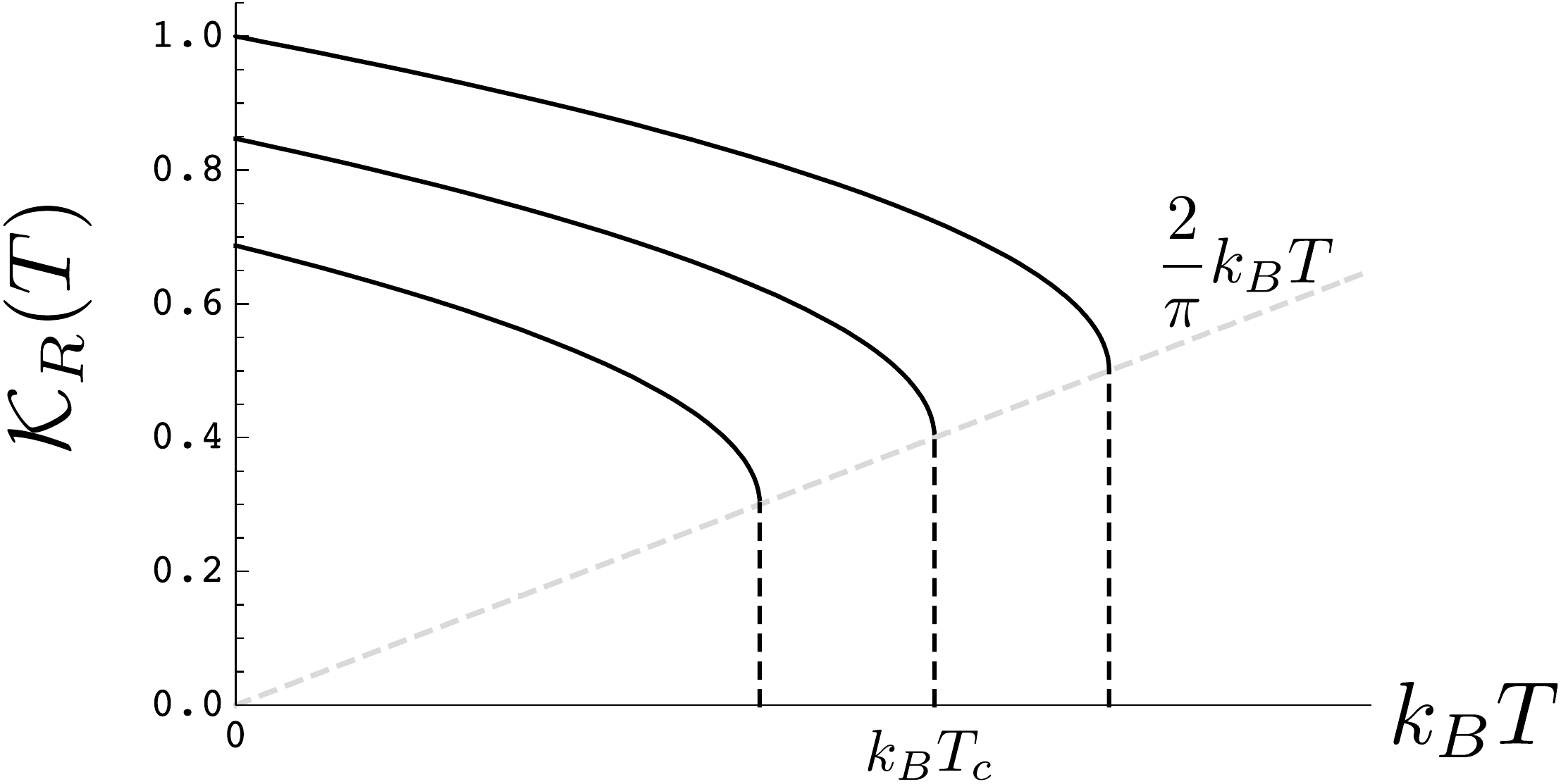}
	\caption{Schematic plot of the (normalized) spin-stiffness ${\cal K}_R$ as a function of temperature. The discontinuity at the transition temperature $T_{{c}}$, represents the onset of a disordered phase of dissociated vortices and anti-vortices pairs. Various curves are plotted to make evident the universality in the jump of spin-stiffness when normalized with the factor $k_B T_{c}$. As $T \to T_{c}^{-}$, ${\cal K}_R = \frac{2}{\pi} k_B T_{c} + c \sqrt{T_{c}-T}$, where $c$ is a nonuniversal positive constant. }
	\label{fig:universaljump}
\end{figure}

We next provide the basics of how to measure ${\cal K}_R(T)$, and thereby detect the universal jump and the associated ''conformality lost transition'' in the recently discovered $2D$ vdW magnets, as the temperature is raised through  $T>T_{{c}}$. We evaluate the spin current-current correlation function ${\cal C}_{\mu\nu}=\langle {j}^s_{\mu}{j}^s_{\nu}\rangle$, where the spin-current is ${j}^s_{\mu}({\bs r})=-J\partial_{\mu}\theta({\bs r})$. The diagonal component, evaluated at the boundary of the magnet, correspond to the correlations between left and right spin-currents (${\cal C}_{yy}$). In momentum space, the spin-current is decomposed into two parts, one part originating with spin-waves and another part with vortices, as follows
\begin{align}\label{eq: sc-correlator}
{\cal C}_{\mu\nu}({\bs q})=\frac{q_{\mu}q_{\nu}}{q^2}\frac{1}{\cal K}+\langle {\cal S}_{q\mu}{\cal S}_{-q\nu} \rangle_v.
\end{align}
Here, ${\cal K}=J/k_B T$ and ${\bs{\cal S}}_q={\cal F}\left[\left(\nabla\theta\right)_v\right]$ is the Fourier transform of the vortex contribution. The second term in the right-hand side in Eq. (\ref{eq: sc-correlator}) may be expressed in terms of vortex correlators $\langle {\cal S}_{q\mu}{\cal S}_{-q\nu} \rangle_v=4\pi^2\left(\delta_{\mu\nu}-q_{\mu}q_{\nu}/q^2\right)\langle n_{q}n_{-q}\rangle/q^2$. Using this result in Eq. (\ref{eq: sc-correlator}), we obtain ${\cal C}_{\mu\nu}({\bs q})=\left(q_{\mu}q_{\nu}/q^2\right)/{\cal K}_R$. The quantity ${\cal K}_R$ represents the renormalized spin-stiffness including the effects of thermally-induced vortices. Thus, ${\cal K}_R$ depends on temperature and obeys the relation given in Eq. (\ref{fig:universaljump}). In the long-wavelength limit, ${\cal K}_R(T)={\cal K}-4\pi^2{\cal K}^2\lim_{q\rightarrow 0}\frac{\langle {n}_{q}{n}_{-q}\rangle}{q^2}$, corresponding to the spin-wave and vortex part, respectively. Since, we are interested in the long-distance behavior of ${\cal K}_R$, we need to evaluate $\langle {n}_{q}{n}_{-q}\rangle$ when $q\rightarrow 0$. We first note that  $\langle {n}_{q}{n}_{-q}\rangle=C_0+C_2q^2+\cdots$, where $C_0$ vanishes by topological-charge neutrality. Thus, the only nonzero contribution in the long-wavelength limit is $C_2$. The evaluation of this coefficient is standard and can be found in Ref. \cite{Kosterlitz2016}. Below, we will relate the spin-current correlation ${\cal C}_{\mu\nu}({\bs q})$ with a measurable quantity, namely the {charge-current cross-correlation. }

{\it Spin-charge conversion}.-- We assume that normal metals have a sufficiently strong spin-orbit coupling to support a considerable SHE. The spin and charge transport in the bulk of  metallic leads are captured by \cite{Tserkovnyak2014}
\begin{align}
{\bf j}^q&\label{eq:chargecurrent}=\frac{\sigma}{e}\nabla\mu_q-\frac{\sigma'}{2e}\nabla\times{\bs \mu}_s,\\
\frac{2e}{\hbar}{\bf j}^s_{n}&\label{eq:spincurrent}=-\frac{\sigma}{2e}\nabla\left(\hat{\bs n}\cdot{\bs \mu}_s\right)-\frac{\sigma'}{e}\left(\hat{\bs n}\times\nabla\right)\mu_q,
\end{align}
where ${\bf j}^q$ and ${\bf j}^s_{n}$ are the charge and spin current (polarized in the $\hat{\bs n}$ direction), respectively. The electrical conductivity is $\sigma$, while $\sigma'$ denotes the spin-Hall conductivity. The spin and charge accumulation, ${\bs \mu}_s$ and $\mu_q$, respectively, are described in the steady-state limit by the equations, $\nabla^2{\mu}_q=0$ and $\nabla^2{\bs \mu}_s={{\bs \mu}_s}/{l^2_s}$, where $l_s$ is the spin-diffusion length in the normal metal. At the metal-magnet interface, the injected spin-current (polarized along $\hat{\bs z}$) is inhomogeneous since it originates from the spins vortex at the magnet. As a result, a spin and charge accumulation, $\mu_s(x,y)$ and $\mu_q(x,y)$, are induced on the normal metal. The bulk equations, Eqs. (\ref{eq:chargecurrent}) and (\ref{eq:spincurrent}), are complemented by the boundary conditions that enforce continuity for the spin-current, ${j}^{s}_{z,y}(x,0)={j}^{s}_{L,z}(x)$ and ${j}^{s}_{z,y}(x,-l)=0$, and charge-currents, $j^q_y(x,0)=j^q_y(x,-l)=0$, at the left lead and where $l$ is the width of the metal. Similar relations holds for the right lead. By simplicity we assume a spin transparent interface, thus the injected spin-current is ${j}^{s}_{L,z}(x)=-J\partial_y\theta(x,y)\left.\right|_{y=0}$. The formal solution for the charge and spin accumulation are written as $\mu_{q,s}(x,y)=\int dx'K_{q,s}(x-x',y){j}^{s}_{L,z}(x')$, with $K_{q}$ and $K_{s}$ the kernels, whose evaluation is detailed in the Supplemental Material. 

{We are interested on the induced charge-current, averaged over the width of the metal, along the $x$-direction at each lead. We obtain $\bar{j}^q_L={2el_s\vartheta}\int^{L_x/2}_{-L_x/2}dx{j}^{s}_{L,z}(x)/{l\hbar L_x}$ in the limit $l\gg l_s$ and small spin Hall angle defined by  $\vartheta=\tan^{-1}\left[\sigma'/\sigma\right]$. Note that the averaged charge-current is independent on the position and proportional to the injected spin-current  averaged  along the interface. Assuming that metals and magnet are in thermal equilibrium, the thermal average of charge- and spin-current are zero. Their correlations, however, which are response functions of the system, will be non-zero and related by  $\langle\bar{j}^q_{L}\bar{j}^q_{R}\rangle=(\xi/L_x)\int^{L_x/2}_{-L_x/2} d\textsc{x}{\cal C}_{yy}(\textsc{x},L_y)$, with $\xi={\pi G_0 l^2_s\vartheta^2\gamma(T)}/{l^2\hbar}$ and $G_0$ the quantum of conductance. The function $\gamma(T)$ is a numerical factor that varies in the range $[4-3.5]$ when $0<T<T_c$.}

{\it Measurement of universal jump in spin-stiffness}.-- In the specific device geometry, the renormalized spin-stiffness can be obtained from the measurement of the {charge-current cross-correlation.} Spin and charge conversion in the metallic leads allow us to relate the spin- and {charge-current correlators}, ${\cal C}_{\mu\nu}$ and ${\cal C}^{(2)}$, respectively. 
The detection of the spin-stiffness can be realized by combined measurements of {current correlations}, since $\text{Tr}\left[{\cal C}_{\mu\nu}({\bs q})\right]=1/{\cal K}_R(T)$. 
The evaluation of Eq. (\ref{eq: sc-correlator}) is carried out in the long-wavelength limit, thus we find that the temperature-dependent spin-stiffness satisfy $1/{\cal K}_R(T)=2\int_{\cal A} d{\bs r}{\cal C}_{yy}({\bs r})$. Here, we have considered axial symmetry and the integration is over the entire area, ${\cal A}$, of the magnet. Although in actual measurements there is only access to the correlations at the boundary of the magnetic sample, this will not restrict the detection of ${\cal K}_R(T)$. 
{In fact, its evaluation can be well approximated by repeating $N$ measurements of $\langle\bar{j}^q_{L}\bar{j}^q_{R}\rangle$ for different lengths $L_y$. This series of measurements correspond to discretize the integration along $y$-direction. To make this approximation efficient, we employ the Gauss--Legendre quadrature method and find,
\begin{align}\label{eq:cross-correlation}
\frac{\xi}{{\cal K}_R(T){\cal A}}={\cal C}^{(2)},
\end{align}
with ${\cal C}^{(2)}=\sum_{i=1}^{N} c_i\langle\bar{j}^q_{L}\bar{j}^q_{R}\rangle_i$ the total charge-current correlator. The measurable correlations in a magnet with length $L^i_y$ is denoted by $\langle\bar{j}^q_{L}\bar{j}^q_{R}\rangle_i$ and $c_i$ are the weights of the approximation, see Supplemental Material for details.} It is expected that only a few measurements will be needed, due to the rapid convergence of this method. Note that to obtain Eq. (\ref{eq:cross-correlation}) we have made two main assumptions. First, the flow of spin-currents across the interface occurs with no resistance, i.e., a large value for the spin mixing conductance. Second, the charge current noise present in metallic leads is not considered. The observation of spin-current fluctuations require a clear mapping to measurable {charge currents}, and hence the presence of current noise might result in additional complications. In practice, these side effects could lead to weaker signals, but we expect them to be less relevant for the cross correlations between the left and right metal, and thus the sharp transition at the critical temperature should not be altered. Details of metal-magnet interface and a realistic treatment of noise are however open issues for the future.

The correlation (\ref{eq:cross-correlation}), which constitute the central result of this Letter, depends inversely on the renormalized spin-stiffness. Thus, we expect that measurements of {resistances fluctuations} can display the temperature-dependent spin-stiffness and the phase transition to a disordered phase of unbound vortices. In Fig. \ref{fig:universaljump} we show the spin-stiffness for various systems characterized by having different transition temperatures. In actual measurements, the universal behaviour of the spin-stiffness at transition temperature will be revealed as the jump occurs along a $2/\pi$-slope straight line. For a specific setup, the jump in the spin-stiffness is proportional to physical parameters related to the actual detector. It is therefore, convenient to normalize the correlation in order to isolate the intrinsic properties of the magnet. 

Although the experimental realization of the device (Fig. \ref{fig:2D-setup}) might be challenging, we consider it as a simple example to illustrate our proposal. A different option, beyond the scope of the present work, would be to consider the pair of metallic leads on top of the magnet. In this case, a spin-accumulation is induced along, and polarized parallel to, the ${z}$-direction. Accordingly, the observation of the spin-stiffness would require the detection of spin-accumulation correlations, which could be done in a spin valve geometry. The latter requires a metallic ferromagnet on top of the leads for a voltage detection.

{\it Summary}.-- We have proposed a setup for measuring the BKT-transition in a 2D magnetic-metal hydrid system. In a non-local geometry, measurements of voltage fluctuations can give access to the temperature dependence of spin-stiffness. In particular, this approach provides direct evidence of the universal jump in the spin-stiffness of the system, a global order parameter for topological order. We hope our proposal will encourage experimental efforts to detect this hallmark of the Berezinski\v{i}-Kosterlitz-Thouless topological phase transition in low-dimensional magnetic systems. 

\begin{acknowledgments}
This work was supported by the Research Council of Norway Project No. 250985 "Fundamentals of Low-dissipative Topological Matter" and the Research Council of Norway through its Centres of Excellence funding scheme, Project No. 262633, "QuSpin".
\end{acknowledgments}

\bibliography{CPT2DvdW}

\begin{thebibliography}{49}%
\makeatletter
\providecommand \@ifxundefined [1]{%
 \@ifx{#1\undefined}
}%
\providecommand \@ifnum [1]{%
 \ifnum #1\expandafter \@firstoftwo
 \else \expandafter \@secondoftwo
 \fi
}%
\providecommand \@ifx [1]{%
 \ifx #1\expandafter \@firstoftwo
 \else \expandafter \@secondoftwo
 \fi
}%
\providecommand \natexlab [1]{#1}%
\providecommand \enquote  [1]{``#1''}%
\providecommand \bibnamefont  [1]{#1}%
\providecommand \bibfnamefont [1]{#1}%
\providecommand \citenamefont [1]{#1}%
\providecommand \href@noop [0]{\@secondoftwo}%
\providecommand \href [0]{\begingroup \@sanitize@url \@href}%
\providecommand \@href[1]{\@@startlink{#1}\@@href}%
\providecommand \@@href[1]{\endgroup#1\@@endlink}%
\providecommand \@sanitize@url [0]{\catcode `\\12\catcode `\$12\catcode
  `\&12\catcode `\#12\catcode `\^12\catcode `\_12\catcode `\%12\relax}%
\providecommand \@@startlink[1]{}%
\providecommand \@@endlink[0]{}%
\providecommand \url  [0]{\begingroup\@sanitize@url \@url }%
\providecommand \@url [1]{\endgroup\@href {#1}{\urlprefix }}%
\providecommand \urlprefix  [0]{URL }%
\providecommand \Eprint [0]{\href }%
\providecommand \doibase [0]{http://dx.doi.org/}%
\providecommand \selectlanguage [0]{\@gobble}%
\providecommand \bibinfo  [0]{\@secondoftwo}%
\providecommand \bibfield  [0]{\@secondoftwo}%
\providecommand \translation [1]{[#1]}%
\providecommand \BibitemOpen [0]{}%
\providecommand \bibitemStop [0]{}%
\providecommand \bibitemNoStop [0]{.\EOS\space}%
\providecommand \EOS [0]{\spacefactor3000\relax}%
\providecommand \BibitemShut  [1]{\csname bibitem#1\endcsname}%
\let\auto@bib@innerbib\@empty
\bibitem [{\citenamefont {Mermin}\ and\ \citenamefont
  {Wagner}(1966)}]{Mermin-Wagner}%
  \BibitemOpen
  \bibfield  {author} {\bibinfo {author} {\bibfnamefont {N.~D.}\ \bibnamefont
  {Mermin}}\ and\ \bibinfo {author} {\bibfnamefont {H.}~\bibnamefont
  {Wagner}},\ }\href {\doibase 10.1103/PhysRevLett.17.1133} {\bibfield
  {journal} {\bibinfo  {journal} {Phys. Rev. Lett.}\ }\textbf {\bibinfo
  {volume} {17}},\ \bibinfo {pages} {1133} (\bibinfo {year}
  {1966})}\BibitemShut {NoStop}%
\bibitem [{\citenamefont {Hohenberg}(1967)}]{Hohenberg1967}%
  \BibitemOpen
  \bibfield  {author} {\bibinfo {author} {\bibfnamefont {P.~C.}\ \bibnamefont
  {Hohenberg}},\ }\href {\doibase 10.1103/PhysRev.158.383} {\bibfield
  {journal} {\bibinfo  {journal} {Phys. Rev.}\ }\textbf {\bibinfo {volume}
  {158}},\ \bibinfo {pages} {383} (\bibinfo {year} {1967})}\BibitemShut
  {NoStop}%
\bibitem [{\citenamefont {Coleman}(1973)}]{Coleman1973}%
  \BibitemOpen
  \bibfield  {author} {\bibinfo {author} {\bibfnamefont {S.}~\bibnamefont
  {Coleman}},\ }\href {\doibase 10.1007/BF01646487} {\bibfield  {journal}
  {\bibinfo  {journal} {Communications in Mathematical Physics}\ }\textbf
  {\bibinfo {volume} {31}},\ \bibinfo {pages} {259} (\bibinfo {year}
  {1973})}\BibitemShut {NoStop}%
\bibitem [{And()}]{Anderson}%
  \BibitemOpen
  \href@noop {} {\bibinfo  {journal} {P. W. Anderson, {\it Basic Notions of
  Condensed Matter Physics} (Benjamin/Cummings, 1984)}\ }\BibitemShut {NoStop}%
\bibitem [{Ber()}]{Berezinskii}%
  \BibitemOpen
\bibfield  {journal} {  }\href@noop {} {\bibinfo  {journal} {V. L. Berezinskii,
  Sov. Phys. JETP 32, 493 (1971)}\ }\BibitemShut {NoStop}%
\bibitem [{\citenamefont {Burch}\ \emph {et~al.}(2018)\citenamefont {Burch},
  \citenamefont {Mandrus},\ and\ \citenamefont {Park}}]{Burch2018}%
  \BibitemOpen
\bibfield  {journal} {  }\bibfield  {author} {\bibinfo {author} {\bibfnamefont
  {K.~S.}\ \bibnamefont {Burch}}, \bibinfo {author} {\bibfnamefont
  {D.}~\bibnamefont {Mandrus}}, \ and\ \bibinfo {author} {\bibfnamefont
  {J.-G.}\ \bibnamefont {Park}},\ }\href {\doibase 10.1038/s41586-018-0631-z}
  {\bibfield  {journal} {\bibinfo  {journal} {Nature}\ }\textbf {\bibinfo
  {volume} {563}},\ \bibinfo {pages} {47} (\bibinfo {year} {2018})}\BibitemShut
  {NoStop}%
\bibitem [{\citenamefont {Kosterlitz}\ and\ \citenamefont
  {Thouless}(1973)}]{Kosterlitz1973}%
  \BibitemOpen
  \bibfield  {author} {\bibinfo {author} {\bibfnamefont {J.~M.}\ \bibnamefont
  {Kosterlitz}}\ and\ \bibinfo {author} {\bibfnamefont {D.~J.}\ \bibnamefont
  {Thouless}},\ }\href {\doibase 10.1088/0022-3719/6/7/010} {\bibfield
  {journal} {\bibinfo  {journal} {Journal of Physics C: Solid State Physics}\
  }\textbf {\bibinfo {volume} {6}},\ \bibinfo {pages} {1181} (\bibinfo {year}
  {1973})}\BibitemShut {NoStop}%
\bibitem [{\citenamefont {Kosterlitz}(1974)}]{Kosterlitz_1974}%
  \BibitemOpen
  \bibfield  {author} {\bibinfo {author} {\bibfnamefont {J.~M.}\ \bibnamefont
  {Kosterlitz}},\ }\href {\doibase 10.1088/0022-3719/7/6/005} {\bibfield
  {journal} {\bibinfo  {journal} {Journal of Physics C: Solid State Physics}\
  }\textbf {\bibinfo {volume} {7}},\ \bibinfo {pages} {1046} (\bibinfo {year}
  {1974})}\BibitemShut {NoStop}%
\bibitem [{\citenamefont {Nelson}\ and\ \citenamefont
  {Kosterlitz}(1977)}]{Nelson_Kosterlitz}%
  \BibitemOpen
  \bibfield  {author} {\bibinfo {author} {\bibfnamefont {D.~R.}\ \bibnamefont
  {Nelson}}\ and\ \bibinfo {author} {\bibfnamefont {J.~M.}\ \bibnamefont
  {Kosterlitz}},\ }\href {\doibase 10.1103/PhysRevLett.39.1201} {\bibfield
  {journal} {\bibinfo  {journal} {Phys. Rev. Lett.}\ }\textbf {\bibinfo
  {volume} {39}},\ \bibinfo {pages} {1201} (\bibinfo {year}
  {1977})}\BibitemShut {NoStop}%
\bibitem [{\citenamefont {Bishop}\ and\ \citenamefont
  {Reppy}(1978)}]{Bishop1978}%
  \BibitemOpen
  \bibfield  {author} {\bibinfo {author} {\bibfnamefont {D.~J.}\ \bibnamefont
  {Bishop}}\ and\ \bibinfo {author} {\bibfnamefont {J.~D.}\ \bibnamefont
  {Reppy}},\ }\href {\doibase 10.1103/PhysRevLett.40.1727} {\bibfield
  {journal} {\bibinfo  {journal} {Phys. Rev. Lett.}\ }\textbf {\bibinfo
  {volume} {40}},\ \bibinfo {pages} {1727} (\bibinfo {year}
  {1978})}\BibitemShut {NoStop}%
\bibitem [{\citenamefont {Gubser}\ and\ \citenamefont
  {Wolf}(1979)}]{Gubser1979}%
  \BibitemOpen
  \bibfield  {author} {\bibinfo {author} {\bibfnamefont {D.}~\bibnamefont
  {Gubser}}\ and\ \bibinfo {author} {\bibfnamefont {S.}~\bibnamefont {Wolf}},\
  }\href {\doibase https://doi.org/10.1016/0038-1098(79)91094-9} {\bibfield
  {journal} {\bibinfo  {journal} {Solid State Communications}\ }\textbf
  {\bibinfo {volume} {32}},\ \bibinfo {pages} {449 } (\bibinfo {year}
  {1979})}\BibitemShut {NoStop}%
\bibitem [{\citenamefont {Hebard}\ and\ \citenamefont
  {Fiory}(1980)}]{Hebard1980}%
  \BibitemOpen
  \bibfield  {author} {\bibinfo {author} {\bibfnamefont {A.~F.}\ \bibnamefont
  {Hebard}}\ and\ \bibinfo {author} {\bibfnamefont {A.~T.}\ \bibnamefont
  {Fiory}},\ }\href {\doibase 10.1103/PhysRevLett.44.291} {\bibfield  {journal}
  {\bibinfo  {journal} {Phys. Rev. Lett.}\ }\textbf {\bibinfo {volume} {44}},\
  \bibinfo {pages} {291} (\bibinfo {year} {1980})}\BibitemShut {NoStop}%
\bibitem [{\citenamefont {Voss}\ \emph {et~al.}(1980)\citenamefont {Voss},
  \citenamefont {Knoedler},\ and\ \citenamefont {Horn}}]{Voss1980}%
  \BibitemOpen
  \bibfield  {author} {\bibinfo {author} {\bibfnamefont {R.~F.}\ \bibnamefont
  {Voss}}, \bibinfo {author} {\bibfnamefont {C.~M.}\ \bibnamefont {Knoedler}},
  \ and\ \bibinfo {author} {\bibfnamefont {P.~M.}\ \bibnamefont {Horn}},\
  }\href {\doibase 10.1103/PhysRevLett.45.1523} {\bibfield  {journal} {\bibinfo
   {journal} {Phys. Rev. Lett.}\ }\textbf {\bibinfo {volume} {45}},\ \bibinfo
  {pages} {1523} (\bibinfo {year} {1980})}\BibitemShut {NoStop}%
\bibitem [{\citenamefont {Wolf}\ \emph {et~al.}(1981)\citenamefont {Wolf},
  \citenamefont {Gubser}, \citenamefont {Fuller}, \citenamefont {Garland},\
  and\ \citenamefont {Newrock}}]{Wolf1981}%
  \BibitemOpen
  \bibfield  {author} {\bibinfo {author} {\bibfnamefont {S.~A.}\ \bibnamefont
  {Wolf}}, \bibinfo {author} {\bibfnamefont {D.~U.}\ \bibnamefont {Gubser}},
  \bibinfo {author} {\bibfnamefont {W.~W.}\ \bibnamefont {Fuller}}, \bibinfo
  {author} {\bibfnamefont {J.~C.}\ \bibnamefont {Garland}}, \ and\ \bibinfo
  {author} {\bibfnamefont {R.~S.}\ \bibnamefont {Newrock}},\ }\href {\doibase
  10.1103/PhysRevLett.47.1071} {\bibfield  {journal} {\bibinfo  {journal}
  {Phys. Rev. Lett.}\ }\textbf {\bibinfo {volume} {47}},\ \bibinfo {pages}
  {1071} (\bibinfo {year} {1981})}\BibitemShut {NoStop}%
\bibitem [{\citenamefont {Epstein}\ \emph {et~al.}(1981)\citenamefont
  {Epstein}, \citenamefont {Goldman},\ and\ \citenamefont
  {Kadin}}]{Epstein1981}%
  \BibitemOpen
  \bibfield  {author} {\bibinfo {author} {\bibfnamefont {K.}~\bibnamefont
  {Epstein}}, \bibinfo {author} {\bibfnamefont {A.~M.}\ \bibnamefont
  {Goldman}}, \ and\ \bibinfo {author} {\bibfnamefont {A.~M.}\ \bibnamefont
  {Kadin}},\ }\href {\doibase 10.1103/PhysRevLett.47.534} {\bibfield  {journal}
  {\bibinfo  {journal} {Phys. Rev. Lett.}\ }\textbf {\bibinfo {volume} {47}},\
  \bibinfo {pages} {534} (\bibinfo {year} {1981})}\BibitemShut {NoStop}%
\bibitem [{\citenamefont {Halperin}\ and\ \citenamefont
  {Nelson}(1978)}]{Halperin1978}%
  \BibitemOpen
  \bibfield  {author} {\bibinfo {author} {\bibfnamefont {B.~I.}\ \bibnamefont
  {Halperin}}\ and\ \bibinfo {author} {\bibfnamefont {D.~R.}\ \bibnamefont
  {Nelson}},\ }\href {\doibase 10.1103/PhysRevLett.41.121} {\bibfield
  {journal} {\bibinfo  {journal} {Phys. Rev. Lett.}\ }\textbf {\bibinfo
  {volume} {41}},\ \bibinfo {pages} {121} (\bibinfo {year} {1978})}\BibitemShut
  {NoStop}%
\bibitem [{\citenamefont {Young}(1979)}]{Young1979}%
  \BibitemOpen
  \bibfield  {author} {\bibinfo {author} {\bibfnamefont {A.~P.}\ \bibnamefont
  {Young}},\ }\href {\doibase 10.1103/PhysRevB.19.1855} {\bibfield  {journal}
  {\bibinfo  {journal} {Phys. Rev. B}\ }\textbf {\bibinfo {volume} {19}},\
  \bibinfo {pages} {1855} (\bibinfo {year} {1979})}\BibitemShut {NoStop}%
\bibitem [{\citenamefont {Zahn}\ \emph {et~al.}(1999)\citenamefont {Zahn},
  \citenamefont {Lenke},\ and\ \citenamefont {Maret}}]{Zahn1999}%
  \BibitemOpen
  \bibfield  {author} {\bibinfo {author} {\bibfnamefont {K.}~\bibnamefont
  {Zahn}}, \bibinfo {author} {\bibfnamefont {R.}~\bibnamefont {Lenke}}, \ and\
  \bibinfo {author} {\bibfnamefont {G.}~\bibnamefont {Maret}},\ }\href
  {\doibase 10.1103/PhysRevLett.82.2721} {\bibfield  {journal} {\bibinfo
  {journal} {Phys. Rev. Lett.}\ }\textbf {\bibinfo {volume} {82}},\ \bibinfo
  {pages} {2721} (\bibinfo {year} {1999})}\BibitemShut {NoStop}%
\bibitem [{\citenamefont {Resnick}\ \emph {et~al.}(1981)\citenamefont
  {Resnick}, \citenamefont {Garland}, \citenamefont {Boyd}, \citenamefont
  {Shoemaker},\ and\ \citenamefont {Newrock}}]{Resnick1981}%
  \BibitemOpen
  \bibfield  {author} {\bibinfo {author} {\bibfnamefont {D.~J.}\ \bibnamefont
  {Resnick}}, \bibinfo {author} {\bibfnamefont {J.~C.}\ \bibnamefont
  {Garland}}, \bibinfo {author} {\bibfnamefont {J.~T.}\ \bibnamefont {Boyd}},
  \bibinfo {author} {\bibfnamefont {S.}~\bibnamefont {Shoemaker}}, \ and\
  \bibinfo {author} {\bibfnamefont {R.~S.}\ \bibnamefont {Newrock}},\ }\href
  {\doibase 10.1103/PhysRevLett.47.1542} {\bibfield  {journal} {\bibinfo
  {journal} {Phys. Rev. Lett.}\ }\textbf {\bibinfo {volume} {47}},\ \bibinfo
  {pages} {1542} (\bibinfo {year} {1981})}\BibitemShut {NoStop}%
\bibitem [{\citenamefont {Voss}\ and\ \citenamefont {Webb}(1982)}]{Voss1982}%
  \BibitemOpen
  \bibfield  {author} {\bibinfo {author} {\bibfnamefont {R.~F.}\ \bibnamefont
  {Voss}}\ and\ \bibinfo {author} {\bibfnamefont {R.~A.}\ \bibnamefont
  {Webb}},\ }\href {\doibase 10.1103/PhysRevB.25.3446} {\bibfield  {journal}
  {\bibinfo  {journal} {Phys. Rev. B}\ }\textbf {\bibinfo {volume} {25}},\
  \bibinfo {pages} {3446} (\bibinfo {year} {1982})}\BibitemShut {NoStop}%
\bibitem [{\citenamefont {Hadzibabic}\ \emph {et~al.}(2006)\citenamefont
  {Hadzibabic}, \citenamefont {Krüger}, \citenamefont {Cheneau}, \citenamefont
  {Battelier},\ and\ \citenamefont {Dalibard}}]{Hadzibabic2006}%
  \BibitemOpen
  \bibfield  {author} {\bibinfo {author} {\bibfnamefont {Z.}~\bibnamefont
  {Hadzibabic}}, \bibinfo {author} {\bibfnamefont {P.}~\bibnamefont {Krüger}},
  \bibinfo {author} {\bibfnamefont {M.}~\bibnamefont {Cheneau}}, \bibinfo
  {author} {\bibfnamefont {B.}~\bibnamefont {Battelier}}, \ and\ \bibinfo
  {author} {\bibfnamefont {J.}~\bibnamefont {Dalibard}},\ }\href {\doibase
  10.1038/nature04851} {\bibfield  {journal} {\bibinfo  {journal} {Nature}\
  }\textbf {\bibinfo {volume} {441}},\ \bibinfo {pages} {1118} (\bibinfo {year}
  {2006})}\BibitemShut {NoStop}%
\bibitem [{\citenamefont {Bedoya-Pinto}\ \emph {et~al.}()\citenamefont
  {Bedoya-Pinto}, \citenamefont {Ji}, \citenamefont {Pandeya}, \citenamefont
  {Gargiani}, \citenamefont {Valvidares}, \citenamefont {Sessi}, \citenamefont
  {Radu}, \citenamefont {Chang},\ and\ \citenamefont
  {Parkin}}]{bedoyapinto2020}%
  \BibitemOpen
  \bibfield  {author} {\bibinfo {author} {\bibfnamefont {A.}~\bibnamefont
  {Bedoya-Pinto}}, \bibinfo {author} {\bibfnamefont {J.-R.}\ \bibnamefont
  {Ji}}, \bibinfo {author} {\bibfnamefont {A.}~\bibnamefont {Pandeya}},
  \bibinfo {author} {\bibfnamefont {P.}~\bibnamefont {Gargiani}}, \bibinfo
  {author} {\bibfnamefont {M.}~\bibnamefont {Valvidares}}, \bibinfo {author}
  {\bibfnamefont {P.}~\bibnamefont {Sessi}}, \bibinfo {author} {\bibfnamefont
  {F.}~\bibnamefont {Radu}}, \bibinfo {author} {\bibfnamefont {K.}~\bibnamefont
  {Chang}}, \ and\ \bibinfo {author} {\bibfnamefont {S.}~\bibnamefont
  {Parkin}},\ }\href@noop {} {}\Eprint {http://arxiv.org/abs/2006.07605}
  {arXiv:2006.07605} \BibitemShut {NoStop}%
\bibitem [{\citenamefont {Novoselov}\ \emph {et~al.}(2004)\citenamefont
  {Novoselov}, \citenamefont {Geim}, \citenamefont {Morozov}, \citenamefont
  {Jiang}, \citenamefont {Zhang}, \citenamefont {Dubonos}, \citenamefont
  {Grigorieva},\ and\ \citenamefont {Firsov}}]{Novoselov2004}%
  \BibitemOpen
  \bibfield  {author} {\bibinfo {author} {\bibfnamefont {K.~S.}\ \bibnamefont
  {Novoselov}}, \bibinfo {author} {\bibfnamefont {A.~K.}\ \bibnamefont {Geim}},
  \bibinfo {author} {\bibfnamefont {S.~V.}\ \bibnamefont {Morozov}}, \bibinfo
  {author} {\bibfnamefont {D.}~\bibnamefont {Jiang}}, \bibinfo {author}
  {\bibfnamefont {Y.}~\bibnamefont {Zhang}}, \bibinfo {author} {\bibfnamefont
  {S.~V.}\ \bibnamefont {Dubonos}}, \bibinfo {author} {\bibfnamefont {I.~V.}\
  \bibnamefont {Grigorieva}}, \ and\ \bibinfo {author} {\bibfnamefont {A.~A.}\
  \bibnamefont {Firsov}},\ }\href {\doibase 10.1126/science.1102896} {\bibfield
   {journal} {\bibinfo  {journal} {Science}\ }\textbf {\bibinfo {volume}
  {306}},\ \bibinfo {pages} {666} (\bibinfo {year} {2004})}\BibitemShut
  {NoStop}%
\bibitem [{\citenamefont {Huang}\ \emph {et~al.}(2017)\citenamefont {Huang},
  \citenamefont {Clark}, \citenamefont {Navarro-Moratalla}, \citenamefont
  {Klein}, \citenamefont {Cheng}, \citenamefont {Seyler}, \citenamefont
  {Zhong}, \citenamefont {Schmidgall}, \citenamefont {McGuire}, \citenamefont
  {Cobden}, \citenamefont {Yao}, \citenamefont {Xiao}, \citenamefont
  {Jarillo-Herrero},\ and\ \citenamefont {Xu}}]{Huang2017}%
  \BibitemOpen
  \bibfield  {author} {\bibinfo {author} {\bibfnamefont {B.}~\bibnamefont
  {Huang}}, \bibinfo {author} {\bibfnamefont {G.}~\bibnamefont {Clark}},
  \bibinfo {author} {\bibfnamefont {E.}~\bibnamefont {Navarro-Moratalla}},
  \bibinfo {author} {\bibfnamefont {D.~R.}\ \bibnamefont {Klein}}, \bibinfo
  {author} {\bibfnamefont {R.}~\bibnamefont {Cheng}}, \bibinfo {author}
  {\bibfnamefont {K.~L.}\ \bibnamefont {Seyler}}, \bibinfo {author}
  {\bibfnamefont {D.}~\bibnamefont {Zhong}}, \bibinfo {author} {\bibfnamefont
  {E.}~\bibnamefont {Schmidgall}}, \bibinfo {author} {\bibfnamefont {M.~A.}\
  \bibnamefont {McGuire}}, \bibinfo {author} {\bibfnamefont {D.~H.}\
  \bibnamefont {Cobden}}, \bibinfo {author} {\bibfnamefont {W.}~\bibnamefont
  {Yao}}, \bibinfo {author} {\bibfnamefont {D.}~\bibnamefont {Xiao}}, \bibinfo
  {author} {\bibfnamefont {P.}~\bibnamefont {Jarillo-Herrero}}, \ and\ \bibinfo
  {author} {\bibfnamefont {X.}~\bibnamefont {Xu}},\ }\href {\doibase
  10.1038/nature22391} {\bibfield  {journal} {\bibinfo  {journal} {Nature}\
  }\textbf {\bibinfo {volume} {546}},\ \bibinfo {pages} {270} (\bibinfo {year}
  {2017})}\BibitemShut {NoStop}%
\bibitem [{\citenamefont {Deng}\ \emph {et~al.}(2018)\citenamefont {Deng},
  \citenamefont {Yu}, \citenamefont {Song}, \citenamefont {Zhang},
  \citenamefont {Wang}, \citenamefont {Sun}, \citenamefont {Yi}, \citenamefont
  {Wu}, \citenamefont {Wu}, \citenamefont {Zhu}, \citenamefont {Wang},
  \citenamefont {Chen},\ and\ \citenamefont {Zhang}}]{Deng2018}%
  \BibitemOpen
  \bibfield  {author} {\bibinfo {author} {\bibfnamefont {Y.}~\bibnamefont
  {Deng}}, \bibinfo {author} {\bibfnamefont {Y.}~\bibnamefont {Yu}}, \bibinfo
  {author} {\bibfnamefont {Y.}~\bibnamefont {Song}}, \bibinfo {author}
  {\bibfnamefont {J.}~\bibnamefont {Zhang}}, \bibinfo {author} {\bibfnamefont
  {N.~Z.}\ \bibnamefont {Wang}}, \bibinfo {author} {\bibfnamefont
  {Z.}~\bibnamefont {Sun}}, \bibinfo {author} {\bibfnamefont {Y.}~\bibnamefont
  {Yi}}, \bibinfo {author} {\bibfnamefont {Y.~Z.}\ \bibnamefont {Wu}}, \bibinfo
  {author} {\bibfnamefont {S.}~\bibnamefont {Wu}}, \bibinfo {author}
  {\bibfnamefont {J.}~\bibnamefont {Zhu}}, \bibinfo {author} {\bibfnamefont
  {J.}~\bibnamefont {Wang}}, \bibinfo {author} {\bibfnamefont {X.~H.}\
  \bibnamefont {Chen}}, \ and\ \bibinfo {author} {\bibfnamefont
  {Y.}~\bibnamefont {Zhang}},\ }\href {\doibase 10.1038/s41586-018-0626-9}
  {\bibfield  {journal} {\bibinfo  {journal} {Nature}\ }\textbf {\bibinfo
  {volume} {563}},\ \bibinfo {pages} {94} (\bibinfo {year} {2018})}\BibitemShut
  {NoStop}%
\bibitem [{\citenamefont {Klein}\ \emph {et~al.}(2018)\citenamefont {Klein},
  \citenamefont {MacNeill}, \citenamefont {Lado}, \citenamefont {Soriano},
  \citenamefont {Navarro-Moratalla}, \citenamefont {Watanabe}, \citenamefont
  {Taniguchi}, \citenamefont {Manni}, \citenamefont {Canfield}, \citenamefont
  {Fern{\'a}ndez-Rossier},\ and\ \citenamefont {Jarillo-Herrero}}]{Klein2018}%
  \BibitemOpen
  \bibfield  {author} {\bibinfo {author} {\bibfnamefont {D.~R.}\ \bibnamefont
  {Klein}}, \bibinfo {author} {\bibfnamefont {D.}~\bibnamefont {MacNeill}},
  \bibinfo {author} {\bibfnamefont {J.~L.}\ \bibnamefont {Lado}}, \bibinfo
  {author} {\bibfnamefont {D.}~\bibnamefont {Soriano}}, \bibinfo {author}
  {\bibfnamefont {E.}~\bibnamefont {Navarro-Moratalla}}, \bibinfo {author}
  {\bibfnamefont {K.}~\bibnamefont {Watanabe}}, \bibinfo {author}
  {\bibfnamefont {T.}~\bibnamefont {Taniguchi}}, \bibinfo {author}
  {\bibfnamefont {S.}~\bibnamefont {Manni}}, \bibinfo {author} {\bibfnamefont
  {P.}~\bibnamefont {Canfield}}, \bibinfo {author} {\bibfnamefont
  {J.}~\bibnamefont {Fern{\'a}ndez-Rossier}}, \ and\ \bibinfo {author}
  {\bibfnamefont {P.}~\bibnamefont {Jarillo-Herrero}},\ }\href {\doibase
  10.1126/science.aar3617} {\bibfield  {journal} {\bibinfo  {journal}
  {Science}\ }\textbf {\bibinfo {volume} {360}},\ \bibinfo {pages} {1218}
  (\bibinfo {year} {2018})}\BibitemShut {NoStop}%
\bibitem [{\citenamefont {Bonilla}\ \emph {et~al.}(2018)\citenamefont
  {Bonilla}, \citenamefont {Kolekar}, \citenamefont {Ma}, \citenamefont {Diaz},
  \citenamefont {Kalappattil}, \citenamefont {Das}, \citenamefont {Eggers},
  \citenamefont {Gutierrez}, \citenamefont {Phan},\ and\ \citenamefont
  {Batzill}}]{Bonilla2018}%
  \BibitemOpen
  \bibfield  {author} {\bibinfo {author} {\bibfnamefont {M.}~\bibnamefont
  {Bonilla}}, \bibinfo {author} {\bibfnamefont {S.}~\bibnamefont {Kolekar}},
  \bibinfo {author} {\bibfnamefont {Y.}~\bibnamefont {Ma}}, \bibinfo {author}
  {\bibfnamefont {H.~C.}\ \bibnamefont {Diaz}}, \bibinfo {author}
  {\bibfnamefont {V.}~\bibnamefont {Kalappattil}}, \bibinfo {author}
  {\bibfnamefont {R.}~\bibnamefont {Das}}, \bibinfo {author} {\bibfnamefont
  {T.}~\bibnamefont {Eggers}}, \bibinfo {author} {\bibfnamefont {H.~R.}\
  \bibnamefont {Gutierrez}}, \bibinfo {author} {\bibfnamefont {M.-H.}\
  \bibnamefont {Phan}}, \ and\ \bibinfo {author} {\bibfnamefont
  {M.}~\bibnamefont {Batzill}},\ }\href {\doibase 10.1038/s41565-018-0063-9}
  {\bibfield  {journal} {\bibinfo  {journal} {Nature Nanotechnology}\ }\textbf
  {\bibinfo {volume} {13}},\ \bibinfo {pages} {289} (\bibinfo {year}
  {2018})}\BibitemShut {NoStop}%
\bibitem [{\citenamefont {Puthirath~Balan}\ \emph {et~al.}(2018)\citenamefont
  {Puthirath~Balan}, \citenamefont {Radhakrishnan}, \citenamefont {Woellner},
  \citenamefont {Sinha}, \citenamefont {Deng}, \citenamefont {Reyes},
  \citenamefont {Rao}, \citenamefont {Paulose}, \citenamefont {Neupane},
  \citenamefont {Apte}, \citenamefont {Kochat}, \citenamefont {Vajtai},
  \citenamefont {Harutyunyan}, \citenamefont {Chu}, \citenamefont {Costin},
  \citenamefont {Galvao}, \citenamefont {Martí}, \citenamefont {van Aken},
  \citenamefont {Varghese}, \citenamefont {Tiwary}, \citenamefont {Malie Madom
  Ramaswamy~Iyer},\ and\ \citenamefont {Ajayan}}]{Puthirath2018}%
  \BibitemOpen
  \bibfield  {author} {\bibinfo {author} {\bibfnamefont {A.}~\bibnamefont
  {Puthirath~Balan}}, \bibinfo {author} {\bibfnamefont {S.}~\bibnamefont
  {Radhakrishnan}}, \bibinfo {author} {\bibfnamefont {C.~F.}\ \bibnamefont
  {Woellner}}, \bibinfo {author} {\bibfnamefont {S.~K.}\ \bibnamefont {Sinha}},
  \bibinfo {author} {\bibfnamefont {L.}~\bibnamefont {Deng}}, \bibinfo {author}
  {\bibfnamefont {C.~d.~l.}\ \bibnamefont {Reyes}}, \bibinfo {author}
  {\bibfnamefont {B.~M.}\ \bibnamefont {Rao}}, \bibinfo {author} {\bibfnamefont
  {M.}~\bibnamefont {Paulose}}, \bibinfo {author} {\bibfnamefont
  {R.}~\bibnamefont {Neupane}}, \bibinfo {author} {\bibfnamefont
  {A.}~\bibnamefont {Apte}}, \bibinfo {author} {\bibfnamefont {V.}~\bibnamefont
  {Kochat}}, \bibinfo {author} {\bibfnamefont {R.}~\bibnamefont {Vajtai}},
  \bibinfo {author} {\bibfnamefont {A.~R.}\ \bibnamefont {Harutyunyan}},
  \bibinfo {author} {\bibfnamefont {C.-W.}\ \bibnamefont {Chu}}, \bibinfo
  {author} {\bibfnamefont {G.}~\bibnamefont {Costin}}, \bibinfo {author}
  {\bibfnamefont {D.~S.}\ \bibnamefont {Galvao}}, \bibinfo {author}
  {\bibfnamefont {A.~A.}\ \bibnamefont {Martí}}, \bibinfo {author}
  {\bibfnamefont {P.~A.}\ \bibnamefont {van Aken}}, \bibinfo {author}
  {\bibfnamefont {O.~K.}\ \bibnamefont {Varghese}}, \bibinfo {author}
  {\bibfnamefont {C.~S.}\ \bibnamefont {Tiwary}}, \bibinfo {author}
  {\bibfnamefont {A.}~\bibnamefont {Malie Madom Ramaswamy~Iyer}}, \ and\
  \bibinfo {author} {\bibfnamefont {P.~M.}\ \bibnamefont {Ajayan}},\ }\href
  {\doibase 10.1038/s41565-018-0134-y} {\bibfield  {journal} {\bibinfo
  {journal} {Nature Nanotechnology}\ }\textbf {\bibinfo {volume} {13}},\
  \bibinfo {pages} {602} (\bibinfo {year} {2018})}\BibitemShut {NoStop}%
\bibitem [{\citenamefont {Song}\ and\ \citenamefont {Gabor}(2018)}]{Song2018}%
  \BibitemOpen
  \bibfield  {author} {\bibinfo {author} {\bibfnamefont {J.~C.~W.}\
  \bibnamefont {Song}}\ and\ \bibinfo {author} {\bibfnamefont {N.~M.}\
  \bibnamefont {Gabor}},\ }\href {\doibase 10.1038/s41565-018-0294-9}
  {\bibfield  {journal} {\bibinfo  {journal} {Nature Nanotechnology}\ }\textbf
  {\bibinfo {volume} {13}},\ \bibinfo {pages} {986} (\bibinfo {year}
  {2018})}\BibitemShut {NoStop}%
\bibitem [{\citenamefont {Huang}\ \emph {et~al.}(2018)\citenamefont {Huang},
  \citenamefont {Clark}, \citenamefont {Klein}, \citenamefont {MacNeill},
  \citenamefont {Navarro-Moratalla}, \citenamefont {Seyler}, \citenamefont
  {Wilson}, \citenamefont {McGuire}, \citenamefont {Cobden}, \citenamefont
  {Xiao}, \citenamefont {Yao}, \citenamefont {Jarillo-Herrero},\ and\
  \citenamefont {Xu}}]{Huang2018}%
  \BibitemOpen
  \bibfield  {author} {\bibinfo {author} {\bibfnamefont {B.}~\bibnamefont
  {Huang}}, \bibinfo {author} {\bibfnamefont {G.}~\bibnamefont {Clark}},
  \bibinfo {author} {\bibfnamefont {D.~R.}\ \bibnamefont {Klein}}, \bibinfo
  {author} {\bibfnamefont {D.}~\bibnamefont {MacNeill}}, \bibinfo {author}
  {\bibfnamefont {E.}~\bibnamefont {Navarro-Moratalla}}, \bibinfo {author}
  {\bibfnamefont {K.~L.}\ \bibnamefont {Seyler}}, \bibinfo {author}
  {\bibfnamefont {N.}~\bibnamefont {Wilson}}, \bibinfo {author} {\bibfnamefont
  {M.~A.}\ \bibnamefont {McGuire}}, \bibinfo {author} {\bibfnamefont {D.~H.}\
  \bibnamefont {Cobden}}, \bibinfo {author} {\bibfnamefont {D.}~\bibnamefont
  {Xiao}}, \bibinfo {author} {\bibfnamefont {W.}~\bibnamefont {Yao}}, \bibinfo
  {author} {\bibfnamefont {P.}~\bibnamefont {Jarillo-Herrero}}, \ and\ \bibinfo
  {author} {\bibfnamefont {X.}~\bibnamefont {Xu}},\ }\href {\doibase
  10.1038/s41565-018-0121-3} {\bibfield  {journal} {\bibinfo  {journal} {Nature
  Nanotechnology}\ }\textbf {\bibinfo {volume} {13}},\ \bibinfo {pages} {544}
  (\bibinfo {year} {2018})}\BibitemShut {NoStop}%
\bibitem [{\citenamefont {Gibertini}\ \emph {et~al.}(2019)\citenamefont
  {Gibertini}, \citenamefont {Koperski}, \citenamefont {Morpurgo},\ and\
  \citenamefont {Novoselov}}]{Gibertini2019}%
  \BibitemOpen
  \bibfield  {author} {\bibinfo {author} {\bibfnamefont {M.}~\bibnamefont
  {Gibertini}}, \bibinfo {author} {\bibfnamefont {M.}~\bibnamefont {Koperski}},
  \bibinfo {author} {\bibfnamefont {A.~F.}\ \bibnamefont {Morpurgo}}, \ and\
  \bibinfo {author} {\bibfnamefont {K.~S.}\ \bibnamefont {Novoselov}},\ }\href
  {\doibase 10.1038/s41565-019-0438-6} {\bibfield  {journal} {\bibinfo
  {journal} {Nature Nanotechnology}\ }\textbf {\bibinfo {volume} {14}},\
  \bibinfo {pages} {408} (\bibinfo {year} {2019})}\BibitemShut {NoStop}%
\bibitem [{\citenamefont {Goennenwein}\ \emph {et~al.}(2015)\citenamefont
  {Goennenwein}, \citenamefont {Schlitz}, \citenamefont {Pernpeintner},
  \citenamefont {Ganzhorn}, \citenamefont {Althammer}, \citenamefont {Gross},\
  and\ \citenamefont {Huebl}}]{Goennenwein2015}%
  \BibitemOpen
  \bibfield  {author} {\bibinfo {author} {\bibfnamefont {S.~T.~B.}\
  \bibnamefont {Goennenwein}}, \bibinfo {author} {\bibfnamefont
  {R.}~\bibnamefont {Schlitz}}, \bibinfo {author} {\bibfnamefont
  {M.}~\bibnamefont {Pernpeintner}}, \bibinfo {author} {\bibfnamefont
  {K.}~\bibnamefont {Ganzhorn}}, \bibinfo {author} {\bibfnamefont
  {M.}~\bibnamefont {Althammer}}, \bibinfo {author} {\bibfnamefont
  {R.}~\bibnamefont {Gross}}, \ and\ \bibinfo {author} {\bibfnamefont
  {H.}~\bibnamefont {Huebl}},\ }\href {\doibase 10.1063/1.4935074} {\bibfield
  {journal} {\bibinfo  {journal} {Applied Physics Letters}\ }\textbf {\bibinfo
  {volume} {107}},\ \bibinfo {pages} {172405} (\bibinfo {year}
  {2015})}\BibitemShut {NoStop}%
\bibitem [{\citenamefont {Cornelissen}\ \emph {et~al.}(2015)\citenamefont
  {Cornelissen}, \citenamefont {Liu}, \citenamefont {Duine}, \citenamefont
  {Youssef},\ and\ \citenamefont {van Wees}}]{Cornelissen2015}%
  \BibitemOpen
  \bibfield  {author} {\bibinfo {author} {\bibfnamefont {L.~J.}\ \bibnamefont
  {Cornelissen}}, \bibinfo {author} {\bibfnamefont {J.}~\bibnamefont {Liu}},
  \bibinfo {author} {\bibfnamefont {R.~A.}\ \bibnamefont {Duine}}, \bibinfo
  {author} {\bibfnamefont {J.~B.}\ \bibnamefont {Youssef}}, \ and\ \bibinfo
  {author} {\bibfnamefont {B.~J.}\ \bibnamefont {van Wees}},\ }\href {\doibase
  10.1038/nphys3465} {\bibfield  {journal} {\bibinfo  {journal} {Nature
  Physics}\ }\textbf {\bibinfo {volume} {11}},\ \bibinfo {pages} {1022}
  (\bibinfo {year} {2015})}\BibitemShut {NoStop}%
\bibitem [{\citenamefont {Lebrun}\ \emph {et~al.}(2018)\citenamefont {Lebrun},
  \citenamefont {Ross}, \citenamefont {Bender}, \citenamefont {Qaiumzadeh},
  \citenamefont {Baldrati}, \citenamefont {Cramer}, \citenamefont {Brataas},
  \citenamefont {Duine},\ and\ \citenamefont {Kläui}}]{lebrun2018}%
  \BibitemOpen
  \bibfield  {author} {\bibinfo {author} {\bibfnamefont {R.}~\bibnamefont
  {Lebrun}}, \bibinfo {author} {\bibfnamefont {A.}~\bibnamefont {Ross}},
  \bibinfo {author} {\bibfnamefont {S.~A.}\ \bibnamefont {Bender}}, \bibinfo
  {author} {\bibfnamefont {A.}~\bibnamefont {Qaiumzadeh}}, \bibinfo {author}
  {\bibfnamefont {L.}~\bibnamefont {Baldrati}}, \bibinfo {author}
  {\bibfnamefont {J.}~\bibnamefont {Cramer}}, \bibinfo {author} {\bibfnamefont
  {A.}~\bibnamefont {Brataas}}, \bibinfo {author} {\bibfnamefont {R.~A.}\
  \bibnamefont {Duine}}, \ and\ \bibinfo {author} {\bibfnamefont
  {M.}~\bibnamefont {Kläui}},\ }\href {\doibase 10.1038/s41586-018-0490-7}
  {\bibfield  {journal} {\bibinfo  {journal} {Nature}\ }\textbf {\bibinfo
  {volume} {561}},\ \bibinfo {pages} {222} (\bibinfo {year}
  {2018})}\BibitemShut {NoStop}%
\bibitem [{\citenamefont {Kwon~Kim}\ and\ \citenamefont
  {Bum~Chung}()}]{Kim2020}%
  \BibitemOpen
  \bibfield  {author} {\bibinfo {author} {\bibfnamefont {S.}~\bibnamefont
  {Kwon~Kim}}\ and\ \bibinfo {author} {\bibfnamefont {S.}~\bibnamefont
  {Bum~Chung}},\ }\href {\doibase arXiv:2003.08956v1} {\
  arXiv:2003.08956v1}\BibitemShut {NoStop}%
\bibitem [{\citenamefont {Sinova}\ \emph {et~al.}(2015)\citenamefont {Sinova},
  \citenamefont {Valenzuela}, \citenamefont {Wunderlich}, \citenamefont
  {Back},\ and\ \citenamefont {Jungwirth}}]{Sinova2015}%
  \BibitemOpen
  \bibfield  {author} {\bibinfo {author} {\bibfnamefont {J.}~\bibnamefont
  {Sinova}}, \bibinfo {author} {\bibfnamefont {S.~O.}\ \bibnamefont
  {Valenzuela}}, \bibinfo {author} {\bibfnamefont {J.}~\bibnamefont
  {Wunderlich}}, \bibinfo {author} {\bibfnamefont {C.~H.}\ \bibnamefont
  {Back}}, \ and\ \bibinfo {author} {\bibfnamefont {T.}~\bibnamefont
  {Jungwirth}},\ }\href {\doibase 10.1103/RevModPhys.87.1213} {\bibfield
  {journal} {\bibinfo  {journal} {Rev. Mod. Phys.}\ }\textbf {\bibinfo {volume}
  {87}},\ \bibinfo {pages} {1213} (\bibinfo {year} {2015})}\BibitemShut
  {NoStop}%
\bibitem [{\citenamefont {Tserkovnyak}\ \emph {et~al.}(2005)\citenamefont
  {Tserkovnyak}, \citenamefont {Brataas}, \citenamefont {Bauer},\ and\
  \citenamefont {Halperin}}]{Tserkovnyak2005}%
  \BibitemOpen
  \bibfield  {author} {\bibinfo {author} {\bibfnamefont {Y.}~\bibnamefont
  {Tserkovnyak}}, \bibinfo {author} {\bibfnamefont {A.}~\bibnamefont
  {Brataas}}, \bibinfo {author} {\bibfnamefont {G.~E.~W.}\ \bibnamefont
  {Bauer}}, \ and\ \bibinfo {author} {\bibfnamefont {B.~I.}\ \bibnamefont
  {Halperin}},\ }\href {\doibase 10.1103/RevModPhys.77.1375} {\bibfield
  {journal} {\bibinfo  {journal} {Rev. Mod. Phys.}\ }\textbf {\bibinfo {volume}
  {77}},\ \bibinfo {pages} {1375} (\bibinfo {year} {2005})}\BibitemShut
  {NoStop}%
\bibitem [{\citenamefont {Oyanagi}\ \emph {et~al.}(2019)\citenamefont
  {Oyanagi}, \citenamefont {Takahashi}, \citenamefont {Cornelissen},
  \citenamefont {Shan}, \citenamefont {Daimon}, \citenamefont {Kikkawa},
  \citenamefont {Bauer}, \citenamefont {van Wees},\ and\ \citenamefont
  {Saitoh}}]{Oyanagi2019}%
  \BibitemOpen
  \bibfield  {author} {\bibinfo {author} {\bibfnamefont {K.}~\bibnamefont
  {Oyanagi}}, \bibinfo {author} {\bibfnamefont {S.}~\bibnamefont {Takahashi}},
  \bibinfo {author} {\bibfnamefont {L.~J.}\ \bibnamefont {Cornelissen}},
  \bibinfo {author} {\bibfnamefont {J.}~\bibnamefont {Shan}}, \bibinfo {author}
  {\bibfnamefont {S.}~\bibnamefont {Daimon}}, \bibinfo {author} {\bibfnamefont
  {T.}~\bibnamefont {Kikkawa}}, \bibinfo {author} {\bibfnamefont {G.~E.~W.}\
  \bibnamefont {Bauer}}, \bibinfo {author} {\bibfnamefont {B.~J.}\ \bibnamefont
  {van Wees}}, \ and\ \bibinfo {author} {\bibfnamefont {E.}~\bibnamefont
  {Saitoh}},\ }\href {\doibase 10.1038/s41467-019-12749-7} {\bibfield
  {journal} {\bibinfo  {journal} {Nature Communications}\ }\textbf {\bibinfo
  {volume} {10}},\ \bibinfo {pages} {4740} (\bibinfo {year}
  {2019})}\BibitemShut {NoStop}%
\bibitem [{\citenamefont {Cornelissen}\ \emph {et~al.}(2017)\citenamefont
  {Cornelissen}, \citenamefont {Oyanagi}, \citenamefont {Kikkawa},
  \citenamefont {Qiu}, \citenamefont {Kuschel}, \citenamefont {Bauer},
  \citenamefont {van Wees},\ and\ \citenamefont {Saitoh}}]{Cornelissen2017}%
  \BibitemOpen
  \bibfield  {author} {\bibinfo {author} {\bibfnamefont {L.~J.}\ \bibnamefont
  {Cornelissen}}, \bibinfo {author} {\bibfnamefont {K.}~\bibnamefont
  {Oyanagi}}, \bibinfo {author} {\bibfnamefont {T.}~\bibnamefont {Kikkawa}},
  \bibinfo {author} {\bibfnamefont {Z.}~\bibnamefont {Qiu}}, \bibinfo {author}
  {\bibfnamefont {T.}~\bibnamefont {Kuschel}}, \bibinfo {author} {\bibfnamefont
  {G.~E.~W.}\ \bibnamefont {Bauer}}, \bibinfo {author} {\bibfnamefont {B.~J.}\
  \bibnamefont {van Wees}}, \ and\ \bibinfo {author} {\bibfnamefont
  {E.}~\bibnamefont {Saitoh}},\ }\href {\doibase 10.1103/PhysRevB.96.104441}
  {\bibfield  {journal} {\bibinfo  {journal} {Phys. Rev. B}\ }\textbf {\bibinfo
  {volume} {96}},\ \bibinfo {pages} {104441} (\bibinfo {year}
  {2017})}\BibitemShut {NoStop}%
\bibitem [{\citenamefont {Ulloa}\ \emph {et~al.}(2019)\citenamefont {Ulloa},
  \citenamefont {Tomadin}, \citenamefont {Shan}, \citenamefont {Polini},
  \citenamefont {van Wees},\ and\ \citenamefont {Duine}}]{Ulloa2019}%
  \BibitemOpen
  \bibfield  {author} {\bibinfo {author} {\bibfnamefont {C.}~\bibnamefont
  {Ulloa}}, \bibinfo {author} {\bibfnamefont {A.}~\bibnamefont {Tomadin}},
  \bibinfo {author} {\bibfnamefont {J.}~\bibnamefont {Shan}}, \bibinfo {author}
  {\bibfnamefont {M.}~\bibnamefont {Polini}}, \bibinfo {author} {\bibfnamefont
  {B.~J.}\ \bibnamefont {van Wees}}, \ and\ \bibinfo {author} {\bibfnamefont
  {R.~A.}\ \bibnamefont {Duine}},\ }\href {\doibase
  10.1103/PhysRevLett.123.117203} {\bibfield  {journal} {\bibinfo  {journal}
  {Phys. Rev. Lett.}\ }\textbf {\bibinfo {volume} {123}},\ \bibinfo {pages}
  {117203} (\bibinfo {year} {2019})}\BibitemShut {NoStop}%
\bibitem [{\citenamefont {Bender}\ \emph {et~al.}(2019)\citenamefont {Bender},
  \citenamefont {Kamra}, \citenamefont {Belzig},\ and\ \citenamefont
  {Duine}}]{Bender2019}%
  \BibitemOpen
  \bibfield  {author} {\bibinfo {author} {\bibfnamefont {S.~A.}\ \bibnamefont
  {Bender}}, \bibinfo {author} {\bibfnamefont {A.}~\bibnamefont {Kamra}},
  \bibinfo {author} {\bibfnamefont {W.}~\bibnamefont {Belzig}}, \ and\ \bibinfo
  {author} {\bibfnamefont {R.~A.}\ \bibnamefont {Duine}},\ }\href {\doibase
  10.1103/PhysRevLett.122.187701} {\bibfield  {journal} {\bibinfo  {journal}
  {Phys. Rev. Lett.}\ }\textbf {\bibinfo {volume} {122}},\ \bibinfo {pages}
  {187701} (\bibinfo {year} {2019})}\BibitemShut {NoStop}%
\bibitem [{\citenamefont {Ando}\ \emph {et~al.}(2011)\citenamefont {Ando},
  \citenamefont {Takahashi}, \citenamefont {Ieda}, \citenamefont {Kajiwara},
  \citenamefont {Nakayama}, \citenamefont {Yoshino}, \citenamefont {Harii},
  \citenamefont {Fujikawa}, \citenamefont {Matsuo}, \citenamefont {Maekawa},\
  and\ \citenamefont {Saitoh}}]{Ando2011}%
  \BibitemOpen
  \bibfield  {author} {\bibinfo {author} {\bibfnamefont {K.}~\bibnamefont
  {Ando}}, \bibinfo {author} {\bibfnamefont {S.}~\bibnamefont {Takahashi}},
  \bibinfo {author} {\bibfnamefont {J.}~\bibnamefont {Ieda}}, \bibinfo {author}
  {\bibfnamefont {Y.}~\bibnamefont {Kajiwara}}, \bibinfo {author}
  {\bibfnamefont {H.}~\bibnamefont {Nakayama}}, \bibinfo {author}
  {\bibfnamefont {T.}~\bibnamefont {Yoshino}}, \bibinfo {author} {\bibfnamefont
  {K.}~\bibnamefont {Harii}}, \bibinfo {author} {\bibfnamefont
  {Y.}~\bibnamefont {Fujikawa}}, \bibinfo {author} {\bibfnamefont
  {M.}~\bibnamefont {Matsuo}}, \bibinfo {author} {\bibfnamefont
  {S.}~\bibnamefont {Maekawa}}, \ and\ \bibinfo {author} {\bibfnamefont
  {E.}~\bibnamefont {Saitoh}},\ }\href {\doibase 10.1063/1.3587173} {\bibfield
  {journal} {\bibinfo  {journal} {Journal of Applied Physics}\ }\textbf
  {\bibinfo {volume} {109}},\ \bibinfo {pages} {103913} (\bibinfo {year}
  {2011})}\BibitemShut {NoStop}%
\bibitem [{\citenamefont {Kosterlitz}(2016)}]{Kosterlitz2016}%
  \BibitemOpen
  \bibfield  {author} {\bibinfo {author} {\bibfnamefont {J.~M.}\ \bibnamefont
  {Kosterlitz}},\ }\href {\doibase 10.1088/0034-4885/79/2/026001} {\bibfield
  {journal} {\bibinfo  {journal} {Reports on Progress in Physics}\ }\textbf
  {\bibinfo {volume} {79}},\ \bibinfo {pages} {026001} (\bibinfo {year}
  {2016})}\BibitemShut {NoStop}%
\bibitem [{\citenamefont {Nelson}(1979)}]{NELSON1979}%
  \BibitemOpen
  \bibfield  {author} {\bibinfo {author} {\bibfnamefont {D.~R.}\ \bibnamefont
  {Nelson}},\ }\href {\doibase https://doi.org/10.1016/0370-1573(79)90116-9}
  {\bibfield  {journal} {\bibinfo  {journal} {Physics Reports}\ }\textbf
  {\bibinfo {volume} {49}},\ \bibinfo {pages} {255 } (\bibinfo {year}
  {1979})}\BibitemShut {NoStop}%
\bibitem [{Cha()}]{Chaikin1995}%
  \BibitemOpen
  \href@noop {} {\bibinfo  {journal} {P. M. Chaikin and T. C. Lubensky, {\it
  Principles of Condensed Matter Physics. Cambridge} (Cambridge University
  Press, 1995)}\ }\BibitemShut {NoStop}%
\bibitem [{Nak()}]{Nakahara}%
  \BibitemOpen
\bibfield  {journal} {  }\href@noop {} {\bibinfo  {journal} {Mikio Nakahara,
  {\it Geometry, Topology and Physics} (CRC Press, 2003)}\ }\BibitemShut
  {NoStop}%
\bibitem [{\citenamefont {Kaplan}\ \emph {et~al.}(2009)\citenamefont {Kaplan},
  \citenamefont {Lee}, \citenamefont {Son},\ and\ \citenamefont
  {Stephanov}}]{PhysRevD.80.125005}%
  \BibitemOpen
\bibfield  {journal} {  }\bibfield  {author} {\bibinfo {author} {\bibfnamefont
  {D.~B.}\ \bibnamefont {Kaplan}}, \bibinfo {author} {\bibfnamefont {J.-W.}\
  \bibnamefont {Lee}}, \bibinfo {author} {\bibfnamefont {D.~T.}\ \bibnamefont
  {Son}}, \ and\ \bibinfo {author} {\bibfnamefont {M.~A.}\ \bibnamefont
  {Stephanov}},\ }\href {\doibase 10.1103/PhysRevD.80.125005} {\bibfield
  {journal} {\bibinfo  {journal} {Phys. Rev. D}\ }\textbf {\bibinfo {volume}
  {80}},\ \bibinfo {pages} {125005} (\bibinfo {year} {2009})}\BibitemShut
  {NoStop}%
\bibitem [{\citenamefont {Nogueira}\ \emph {et~al.}(2019)\citenamefont
  {Nogueira}, \citenamefont {van~den Brink},\ and\ \citenamefont
  {Sudb\o{}}}]{PhysRevD.100.085005}%
  \BibitemOpen
  \bibfield  {author} {\bibinfo {author} {\bibfnamefont {F.~S.}\ \bibnamefont
  {Nogueira}}, \bibinfo {author} {\bibfnamefont {J.}~\bibnamefont {van~den
  Brink}}, \ and\ \bibinfo {author} {\bibfnamefont {A.}~\bibnamefont
  {Sudb\o{}}},\ }\href {\doibase 10.1103/PhysRevD.100.085005} {\bibfield
  {journal} {\bibinfo  {journal} {Phys. Rev. D}\ }\textbf {\bibinfo {volume}
  {100}},\ \bibinfo {pages} {085005} (\bibinfo {year} {2019})}\BibitemShut
  {NoStop}%
\bibitem [{\citenamefont {Tserkovnyak}\ and\ \citenamefont
  {Bender}(2014)}]{Tserkovnyak2014}%
  \BibitemOpen
  \bibfield  {author} {\bibinfo {author} {\bibfnamefont {Y.}~\bibnamefont
  {Tserkovnyak}}\ and\ \bibinfo {author} {\bibfnamefont {S.~A.}\ \bibnamefont
  {Bender}},\ }\href {\doibase 10.1103/PhysRevB.90.014428} {\bibfield
  {journal} {\bibinfo  {journal} {Phys. Rev. B}\ }\textbf {\bibinfo {volume}
  {90}},\ \bibinfo {pages} {014428} (\bibinfo {year} {2014})}\BibitemShut
  {NoStop}%
\end{thebibliography}%

\clearpage
\onecolumngrid
\section{Supplemental Material}

In this Supplemental Material, we show explicitly the conversion (via SHE and ISHE) from spin- to charge-currents. Particularly, we focus on the mapping of the spin-current correlator and the measurable charge-current fluctuations in the proposed device geometry.
\subsection{Boundary Conditions}
Assuming the spin accumulation is along $\hat{\bs z}$, the spin current at the left and right interfaces is
\begin{align}
j^{s}_{L,z}=\frac{g^{\uparrow\downarrow}}{4\pi}\left(\mu_{sL}-\hbar\dot{\theta}\right),\\
j^{s}_{R,z}=\frac{g^{\uparrow\downarrow}}{4\pi}\left(\mu_{sR}-\hbar\dot{\theta}\right).
\end{align}

In addition, the spin-current in the bulk of the magnet is polarized along $\hat{\bs z}$ and given by ${\bs j}^s({\bs r})=-{J}\nabla\theta({\bs r})$. For all points at the boundary it satisfies
\begin{align}
\frac{g^{\uparrow\downarrow}}{4\pi}\left(\mu_{sL}-\hbar\dot{\theta}\right)&=\left.-J\frac{\partial\theta({\bs r})}{\partial y}\right|_{y=0},\\
\frac{g^{\uparrow\downarrow}}{4\pi}\left(\mu_{sR}-\hbar\dot{\theta}\right)&=\left.-J\frac{\partial\theta({\bs r})}{\partial y}\right|_{y=L_y}.
\end{align}
Henceforth, we assume that net spin-currents, ${\bf j}^{s}_{L,z}$ and ${\bf j}^{s}_{R,z}$, flow with no resistance at the interfaces. This imply $j^{s}_{L,z}=-J\left.{\partial_y\theta({\bs r})}\right|_{y=0}$ and $j^{s}_{R,z}=-J\left.{\partial_y\theta({\bs r})}\right|_{y=L_y}$.

\subsection{Spin and charge conversion}
The spin and charge transport in the bulk of the metallic leads are governed by 
\begin{align}
{\bf j}^q&\label{eq: sct1}=\frac{\sigma}{e}\nabla\mu_q-\frac{\sigma'}{2e}\nabla\times{\bs \mu}_s,\\
\frac{2e}{\hbar}{\bf j}^s_{n}&\label{eq: sct2}=-\frac{\sigma}{2e}\nabla\left(\hat{\bs n}\cdot{\bs \mu}_s\right)-\frac{\sigma'}{e}\left(\hat{\bs n}\times\nabla\right)\mu_q,
\end{align}
where ${\bf j}^q$ and ${\bf j}^s_{n}$ are the charge-current and spin-current, respectively. The unit vector $\hat{\bs n}$ represents the direction of spin polarization, which in the present case is assumed to be along the $z$ direction.  The electrical conductivity is $\sigma$, while $\sigma'$ stands for the spin Hall conductivity. The spin accumulation and electrochemical potential are respectively, ${\bs \mu}_s$ and $\mu_q$, which in the steady-state limit are described by the equations,
\begin{align}
\nabla^2{\mu}_q&\label{eq:decharge}=0,\\
\nabla^2{\bs \mu}_s&\label{eq:despin}=\frac{{\bs \mu}_s}{l^2_s},
\end{align}
where $l_s$ is the spin-diffusion length in the NM. According with the device geometry in Fig. \ref{fig:2D-setup}, the spin-current propagates in the XY-plane with an out-of-plane spin-polarization. At the interface, the injected spin-current is inhomogeneous due to the presence of spin textures (vortices) at the magnet. Thus, the induced spin and charge accumulation on the normal metal, $\mu_s(x,y)$ and $\mu_q(x,y)$, needs to be determined. 

The set of equations Eqs. (\ref{eq: sct1}-\ref{eq:despin}) are complemented by the boundary conditions (BC) that enforce continuity for the spin- and charge-current at the boundaries of the metallic leads. For instance, at the left lead, the BC for the spin-current are given by 
\begin{subequations}
\begin{align}
{j}^{s}_{z,y}(x,y)\left.\right|_{y=0}&\label{eq:BCspin1}={j}^{s}_{L,z}(x),\\
{j}^{s}_{z,y}(x,y)\left.\right|_{y=-l}&\label{eq:BCspin2}=0,
\end{align}
\end{subequations}
and for the charge-current
\begin{subequations}
\begin{align}
j^q_y(x,y)\left.\right|_{y=0}&\label{eq:BCcharge1}=0,\\
j^q_y(x,y)\left.\right|_{y=-l}&\label{eq:BCcharge2}=0,
\end{align}
\end{subequations}
where $l$ and $L_x$ are the width and length of the normal metal. Similar BC are established for the right lead. 

The formal solution of Eqs. (\ref{eq:decharge}) and (\ref{eq:despin}) for the charge and spin accumulation are 
\begin{align}
\mu_q(x,y)&=\int dx'K_q(x-x',y){j}^{s}_{L,z}(x'),\\
\mu_s(x,y)&=\int dx'K_s(x-x',y){j}^{s}_{L,z}(x'),
\end{align} 
with $K_{q}$ and $K_{s}$ the kernels. The charge and spin accumulation are explicitly written in terms of the injected spin current at the metal-magnet boundary, i.e., ${j}^{s}_{L,z}(x)=-J\partial_y\theta({\bs r})\left.\right|_{y=0}$. To find the kernels we Fourier transform along $x$-direction and replace them in Eqs. (\ref{eq:decharge}) and (\ref{eq:despin}) to obtain,
\begin{align}
\left(\frac{\partial^2}{\partial y^2}
-k^2_x\right)\bar{K}_q(k_x,y)&\label{eq:kbq}=0,\\
\left(\frac{\partial^2}{\partial y^2}-\frac{1}{l^2_s}-k^2_x\right)\bar{K}_s(k_x,y)&=0,
\end{align}
with $\bar{K}_{q,s}(k_x,y)=\int dx K_{q,s}(x,y) e^{-ik_xx}$. 
The general solution for $\bar{K}_{s}$ and $\bar{K}_{q}$ are therefore, $\bar{K}_s(k_x,y)=A e^{y/\alpha}+B e^{-y/\alpha}$ and $\bar{K}_q(k_x,y)=C e^{y k_x}+D e^{-y k_x}$,
where $1/\alpha^2=k^2_x+1/l^2_s$.
The set of integration constants, $A, B, C$ and $D$, are obtained from the boundary conditions, Eqs. (\ref{eq:BCspin1}-\ref{eq:BCcharge2}), and
satisfy the algebraic equations,
\begin{align}
0&\label{eq:ae1}=\frac{2e^2}{\hbar}+\frac{\sigma}{2\alpha} \left(A-B\right)+ik_x\sigma'\left(C+D \right),\\
0&\label{eq:ae2}=\frac{\sigma}{2\alpha} \left({Ae^{-l/\alpha}-Be^{l/\alpha}}\right)+ik_x\sigma'\left(Ce^{-l k_x}+De^{l k_x}\right),\\
0&\label{eq:ae3}={k_x}{\sigma}\left(C-D\right)+\frac{ik_x\sigma'}{2}\left({A +B}\right),\\
0&\label{eq:ae4}={k_x}\sigma\left(C e^{-l k_x}-D e^{l k_x}\right)+\frac{ik_x\sigma'}{2}\left({A e^{-l/\alpha}+B e^{l/\alpha}}\right).
\end{align}

A similar procedure apply for the right lead. Note that the coefficients $C$ and $D$ are undetermined when $k_x=0$ and that Eqs. (\ref{eq:ae3}) and (\ref{eq:ae4}) are relevant to determine $C$ and $D$ only when $k_x\neq 0$. Thus, we look for solutions with the form $A_{k}=\frac{2\pi}{L_x}A_0\delta(k)+\tilde{A}_{k}$, with $\tilde{A}_k$ being some analytic function satisfying Eqs. (\ref{eq:ae1})-(\ref{eq:ae4}). Similar considerations are done for the rest of coefficients $B, C$ and $D$.

The solution for the kernels can be written as $\bar{K}_s(k_x,y)=2\pi\bar{K}^0_s(y)\delta(k_x)/L_x+\tilde{K}_s(k_x,y)$, with $\bar{K}^0_{s}(y)$ and $\tilde{K}_{s}(0,y)$ given by
\begin{align}
\bar{K}^0_{s}(y)&=-\frac{4l_s e^2}{\sigma\hbar }\frac{\cosh\left[\frac{l+y}{l_s}\right]}{\sinh\left[\frac{l}{l_s}\right]},\\
\tilde{K}_{s}(k_x,y)&=-\frac{4\alpha e^2}{\sigma\hbar}\frac{\cosh\left[\frac{l+y}{\alpha}\right]}{\sinh\left[\frac{l}{\alpha}\right]},
\end{align}

while for the charge kernel $\bar{K}_q(k_x,y)=2\pi\bar{K}^0_q(y)\delta(k_x)/L_x+\tilde{K}_q(k_x,y)$, with
\begin{align}
\bar{K}_{q}(k_x,y)&=\frac{2ie^2\alpha\sigma'}{\sigma^2\hbar}\frac{\cosh\left[(l+y)k_x\right]\coth\left[\frac{l}{\alpha}\right]-\cosh\left[y k_x\right]\text{csch}\left[\frac{l}{\alpha}\right]}{\sinh\left[lk_x\right]},
\end{align}
in the limit $\sigma'\ll\sigma$. The factor $\bar{K}^0_{q}(y)$ is undetermined, since it is defined in terms of $C_0$ and $D_0$. Nevertheless, it determines the homogeneous part of the chemical potential, $\mu_q(x,y)=\mu_0+\tilde{\mu}_q(x,y)$, with $\mu_0=(C_0+D_0)\int dx'{j}^{s}_{L,z}(x')/{L_x}$. Therefore $\bar{K}^0_{q}(y)$ is not relevant to obtain the charge-current.

We now evaluate the charge-current along the $x$-direction, which explicitly reads in terms of the kernels as,
\begin{align}
j^q_x(x,y)=\int dx'\int dk_x\left[\frac{\sigma}{e}ik_x\bar{K}_q(k_x,y)-\frac{\sigma'}{2e}\partial_y\bar{K}_s(k_x,y)\right]e^{ik_x(x-x')}{j}^{s}_{L,z}(x').
\end{align}
The average over the width of the lead satisfy,
\begin{align}
\bar{j}^q_x(x)=\frac{1}{l}\int^0_{-l} dyj^q_x(x,y)&\nonumber=-\frac{\pi\sigma'}{eL_x}\int dx'\int \frac{dk_x}{2\pi}\left\{\frac{1}{l}\int^0_{-l} dy \left[\partial_y\bar{K}^0_s(y)\delta(k_x)\right]\right\}e^{ik_x(x-x')}{j}^{s}_{L,z}(x')\\
&\nonumber\qquad\qquad+\int dx'\int \frac{dk_x}{2\pi}\underbrace{\left\{\frac{1}{l}\int^0_{-l} dy \left[\frac{\sigma}{e}ik_x\bar{K}_q(k_x,y)-\frac{\sigma'}{2e}\partial_y\tilde{K}_s(k_x,y)\right]\right\}}_{=0}e^{ik_x(x-x')}{j}^{s}_{L,z}(x')\\
&\label{eq:acq}=\frac{2el_s\sigma'}{l\sigma \hbar}\tanh\left[\frac{l}{2l_s}\right]\frac{1}{L_x}\int dx'{j}^{s}_{L,z}(x').
\end{align}

Note that when the injected spin-current is homogeneous, Eq. (\ref{eq:acq}) reduces to the well known result  $\bar{j}^q_{L}=\left({2el_s\sigma'}/l\sigma\hbar\right)j^s_{L,z}\tanh\left[{l}/{2l_s}\right]$. It is worth commenting that the averaged charge-current $\bar{j}^q_x$ is independent on the position $x$. This is expected from charge conservation, which in the steady state and averaged over the $y$ direction, reads $\partial_x\bar{j}^q_x+\frac{1}{l}\left(j^q_y(x,0)-j^q_y(x,-l)\right)=0$. Considering the boundary conditions for the charge current, we clearly observes that the average current is position independent. 

The expression for the charge-current, Eq. (\ref{eq:acq}), can be simplified as $\bar{j}^q_L=({2el_s\vartheta}/{l\hbar })\bar{j}^{s}_{L,z}$ for $l\gg l_s$ and small spin Hall angle $\vartheta=\tan^{-1}\left[\sigma'/\sigma\right]$, where $\bar{j}^{s}_{L,z}$ corresponds to the average spin-current injected at the left lead. The result for $\bar{j}^q_L$ was obtained for a certain spin configuration that, at the boundary with the metal, induces an inhomogeneous spin-current on the metallic leads. At finite temperature, however, the spin system fluctuates and therefore a thermal average over all possible spin configurations is needed. Thus, the correlation of the charge current in the normal metals and the spin currents in the magnetic insulator are related by
\begin{align}\label{eq:chargecorrelator}
\langle\bar{j}^q_{L}\bar{j}^q_{R}\rangle=\frac{4e^2l^2_s\vartheta^2}{l^2\hbar^2L^2_x}\int^{L_x/2}_{-L_x/2} dx\int^{L_x/2}_{-L_x/2} dx'\langle{j}^{s}_{L,z}(x){j}^{s}_{R,z}(x')\rangle
\end{align}
where $\langle,\rangle$ stands for a statistical average.

To evaluate the correlator, Eq. (\ref{eq:chargecorrelator}), we need to find a useful expression for the double integral ${\cal I}=\int^{L_x/2}_{-L_x/2} \int^{L_x/2}_{-L_x/2} dx dx'\langle {j}^{s}_{L,z}(x){j}^{s}_{R,z}(x')\rangle$. First, we note that the correlation $\langle{j}^{s}_{L,z}(x){j}^{s}_{R,z}(x')\rangle$ depends on the relative distance $\textsc{x}\equiv x-x'$, and thus, we define the function $f(\textsc{x})=\langle{j}^{s}_{L,z}(x){j}^{s}_{R,z}(x')\rangle.$ Next, we discretize the integral ${\cal I}$ on a square array of $N\times N$ points, where $x\rightarrow x_i=-L_x/2+L_x i/N$, with $i=1,\dots,N$, thus
\begin{align}
{\cal I}=\lim_{N\rightarrow\infty} \sum^{N}_{i=1}\sum^{N}_{j=1}\left(\frac{L_x}{N}\right)^2f\left[i-j\right]
\end{align}
where we have introduced the  notation $f\left[i-j\right]=f\left[\frac{L_x}{N}(i-j)\right]$. We expand the previous summation and regroup terms as follows
\begin{align}
\sum^{N}_{i=1}\sum^{N}_{j=1}\left(\frac{L_x}{N}\right)^2f\left[i-j\right] 
=\nonumber\left(\frac{L_x}{N}\right)^2\left(A_0f[0]+A_1(f[-1] + f[1]) +\cdots 
+ A_{\Delta} (f[\Delta]+ f[-\Delta])\right.\\
\left.+\cdots+A_{N-1}(f[N-1] + f[-(N-1)])\right).
\end{align}

The full summation that runs over all the points in a $N\times N$ matrix, is separated by a summation of elements labeled as $-(N-1) \le \Delta\le N-1$. It is then readily seen that the coefficients are $A_{\Delta}=N-|\Delta|$. Considering the previous summation of all the $i,j$ points such that $i=j$, this corresponds to adding the $N$ points on the diagonal, yielding the result $N f[0]$. When $i=j\pm 1$, we repeat the same procedure, obtaining $(N-1)f[\pm 1]$, and so on and so forth. Thus, we obtain the relation
\begin{align}
\sum^{N}_{i=1}\sum^{N}_{j=1}\left(\frac{L_x}{N}\right)^2f\left[i-j\right]&\nonumber=\sum^{N-1}_{\Delta=-(N-1)}\left(\frac{L_x}{N}\right)^2A_{\Delta} f[\Delta]\\
&= \sum^{N-1}_{\Delta=-(N-1)}\left(\frac{L_x}{N}\right)^2(N-|\Delta|) f[\Delta].
\end{align}
Since $f[i-j]=f[j-i]$, we have $f[-\Delta] = f[\Delta]$. Then,
\begin{align}\label{eq:Iresult}
{\cal I}=\lim_{N\rightarrow \infty}
\sum^{N-1}_{\Delta=-(N-1)}\left(\frac{L_x}{N}\right)^2(N-|\Delta|) f[\Delta]
= \int^{L_x}_{-L_x}d\textsc{x}\left( L_x-|\textsc{x}|\right)f(\textsc{x})
=2\int^{L_x}_{0}d\textsc{x}\left( L_x-\textsc{x}\right)f(\textsc{x}).
\end{align}
The function $f(\textsc{x})$ represent the correlation function with a power-law decay at temperatures below the transition temperature $T_c$. This suggests the following approximation to the integral in Eq. (\ref{eq:Iresult}): $f(\textsc{x})$ features fairly slow power-law decay whereas the factor $L_x-\textsc{x}$ varies from $L_x$ to $0$ on the interval $\textsc{x} \in [0,L_x]$. The factor $L_x-\textsc{x}$ is therefore the most rapidly varying. Hence, we will approximate the integral by
\begin{align}
{\cal I}=2\int^{L_x}_{0}d\textsc{x}\left( L_x-\textsc{x}\right)f(\textsc{x}) \approx 2\bar{f} \int^{L_x}_{0}d\textsc{x} (L_x-\textsc{x}),
\end{align}
where $\bar{f} = (1/L_x) \int_0^{L_x} d\textsc{x} f(\textsc{x})$ is the spatial average of $f(\textsc{x})$ on the interval. Thus, we obtain
\begin{align}
{\cal I} \approx L_x \int^{L_x}_{0}d\textsc{x} f({\textsc{x}}),
\end{align}
{Note that the domain of integration now is twice the system size, since the upper limit on the interface-coordinates is $L_x/2$. This is because $\textsc{x}$ describes the relative coordinate $x-x^{\prime}$. By numerical inspection we corroborate that this integral scales with temperature as ${\cal I}[L_x]=\gamma(T){\cal I}[L_x/2]$ for large system size $L_x\gg l_s$. The factor $\gamma(T)\in[4-3.5]$ when $0<T<T_c$ and has a very smooth transition with $L_x$ when $T\rightarrow T_c$}. Thus, the correlator Eq. (\ref{eq:chargecorrelator}) is simply given as,
\begin{align}
\langle\bar{j}^q_{L}\bar{j}^q_{R}\rangle
=\frac{\pi G_0 l^2_s\vartheta^2\gamma(T)}{l^2\hbar L_x}\int^{L_x/2}_{-L_x/2} d\textsc{x}{\cal C}_{yy}(\textsc{x},L_y),
\end{align}
with $G_0$ the quantum of conductance.

\subsection{Spin and Current Correlations Mapping}
We are interested in the evaluation of the spin-current--spin-current correlation function ${\cal C}_{\mu\nu}=\langle {j}^s_{\mu}{j}^s_{\nu}\rangle$, where the spin-current in the bulk of the magnet is ${j}^s_{\mu}({\bs r})=-J\partial_{\mu}\theta({\bs r})$. In the long-wave limit the correlator obeys 
\begin{align}\label{eq:renspinstiff}
\text{Tr}\left[{\cal C}_{\mu\nu}({\bs q})\right]=2\int_{\cal A}d{\bs r}{\cal C}_{yy}({\bs r})=\frac{1}{{\cal K}_R(T)}.
\end{align}
In an actual measurement, we have access to the spin-current correlator only at the interface of the sample, see Eq. (\ref{eq:voltcorrelator}). In addition, the length of the magnet ($L_y$) is fixed. Therefore, to connect Eqs. (\ref{eq:voltcorrelator}) and (\ref{eq:renspinstiff}) the integration along $y$ needs to be approximated. In practice, this corresponds to taking a series of measurements by changing the length $L_y$. We can then use the Gauss–Legendre quadrature method to approximate the integral along the $y$-direction. First, we choose $L^1_y(L^N_y)$ as the lower(upper) width of the magnet and write $\int^{L^N_y}_{L^1_y}d\textsc{y}{\cal C}_{yy}(\textsc{x},\textsc{y})=\left(L^N_y-L^1_y\right)\int^1_{-1}d\zeta {\cal C}_{yy}(\textsc{x},\zeta)/2$, where $\textsc{y}=\left(L^N_y-L^1_y\right)\zeta/2+\left(L^N_y+L^1_y\right)/2$. After discretizing, we obtain
\begin{align}
\frac{1}{{\cal K}_R(T)}\approx \Delta L_y\sum_{i=1}^{N} c_i \int^{L_x/2}_{-L_x/2}d\textsc{x}{\cal C}_{yy}(\textsc{x},{\zeta_i}),
\end{align}
with $\Delta L_y=L^N_y-L^1_y$. The points $\zeta_i$ correspond to the roots of a Legendre polynomial of degree $N$, $P_N(\zeta_i)=0$. The zeros are real and lie on the interval $(-1,1)$; $-1 < \zeta_1<\dots < \zeta_N<1$. The weights in this approximation are defined by  $c_i=2\left((1-\zeta^2_i)P'_N(\zeta_i)\right)^{-2}$. Note that each point $\zeta_i$ is associated to a specific length $L^i_y$. Thus, considering that ${\cal C}_{yy}(\textsc{x},\zeta_i)$ represents the spin-current correlation for a certain length of the sample, we can write 

\begin{align}\label{eq:voltcorrelator}
\frac{\xi}{L_x\Delta L_y{\cal K}_R(T)}\approx{\cal C}^{(2)},
\end{align}
with ${\cal C}^{(2)}=\sum_{i=1}^{N} c_i\langle\bar{j}^q_{L}\bar{j}^q_{R}\rangle_i$ the total voltage correlator, $\langle\bar{j}^q_{L}\bar{j}^q_{R}\rangle_i$ denotes the measurable voltage fluctuations in a system with $L^i_y$ the length of the magnet and $\xi={\pi G_0 l^2_s\vartheta^2\gamma(T)}/{l^2\hbar}$. If $L^N_y\gg L^1_y$, the factor $L_x\Delta L$ can be approximated to the total area ${\cal A}$ of the magnet. From the experimental point of view, only a few measurements will be needed, since the Gauss–Legendre quadrature method converges quickly to the desired result.

\end{document}